\documentclass[twocolumn]{aastex61}

\usepackage{natbib}
\usepackage[caption=false]{subfig}
\usepackage{amsmath,amssymb}

\newcommand{\sha}{KIC 1255 b}
\newcommand{\shStar}{KIC 1255}
\newcommand{\kepler}{{\it Kepler}}

\shorttitle{KIC 12557548 b back to ``normal''}
\shortauthors{Schlawin et al.}

\begin{document}

\title{Back to ``Normal'' for the Disintegrating Planet Candidate KIC 12557548 \MakeLowercase{b}}


\correspondingauthor{E. Schlawin}
\email{eas342@email.arizona.edu}

\author[0000-0001-8291-6490]{Everett Schlawin}
\affiliation{Steward Observatory, Tucson AZ 85721}

\author{Teruyuki Hirano}
\affiliation{Department of Earth and Planetary Sciences, Tokyo Institute of Technology, Tokyo 152-8550, Japan}

\author{Hajime Kawahara}
\affiliation{Department of Earth and Planetary Science, The University of Tokyo, Tokyo 113-0033, Japan}
\affiliation{Research Center for the Early Universe, School of Science, The University of Tokyo, Tokyo 113-0033, Japan}

\author{Johanna Teske}
\altaffiliation{Hubble Fellow, jteske@carnegiescience.edu} 
\affiliation{Observatories of the Carnegie Institution for Science, 813 Santa Barbara Street, Pasadena, CA 91101}

\author{Elizabeth M. Green}
\affiliation{Steward Observatory, Tucson AZ 85721}

\author[0000-0002-3627-1676]{Benjamin V. Rackham}
\affiliation{Steward Observatory, Tucson AZ 85721}
\affiliation{Earths in Other Solar Systems Team, NASA Nexus for Exoplanet System Science.}

\author{Jonathan Fraine}
\affiliation{Space Telescope Science Institute, 3700 San Martin Drive, Baltimore, MD 21218, USA}

\author{Rafia Bushra}
\affiliation{Steward Observatory, Tucson AZ 85721}

\begin{abstract}
KIC 12557548 b is first of a growing class of intriguing disintegrating planet candidates, which lose mass in the form of \added{a metal rich vapor that condenses into} dust particles \deleted{escaping their surface}.
Here, we follow up two perplexing observations of the system: 1) the transits appeared shallower than average in 2013 and 2014 and 2) the parameters derived from a high resolution spectrum of the star differed from other results using photometry and low resolution spectroscopy.
We observe 5 transits of the system with the 61-inch Kuiper telescope in 2016 and show that they are consistent with photometry from the \kepler\ spacecraft in 2009-2013\added{, suggesting that the dusty tail has returned to normal length and mass}.
We also evaluate high resolution archival spectra from the Subaru HDS spectrograph and find them to be consistent with a main-sequence T$_\mathrm{eff}$=4440 $\pm$70 K star in agreement with the photometry and low resolution spectroscopy.
This disfavors the hypothesis that planet disintegration affected the analysis of prior high resolution spectra of this star.
We apply Principal Component Analysis to the \kepler\ long cadence data to understand the modes of disintegration.
There is a tentative 491 day periodicity of the second principal component, which corresponds to possible long-term evolution of the dust grain sizes, though the mechanism on such long timescales remains unclear.
\end{abstract}



\section{Introduction}\label{sec:intro}
\citet{rappaport} first reported the discovery of a highly unusual system in the \kepler\ field, KIC 12557548.
Its light curve exhibits flux dips with a period of \replaced{0.653}{0.654} days.
Surprisingly, the flux dips are neither symmetric in shape nor constant in amplitude, varying from $\sim$0\% to $\sim$1.2\%.
The amplitudes of the flux dips are highly stochastic and unpredictable from one event to the next \replaced{0.653}{0.654} days later.
This is \replaced{radially}{radically} different from a transiting planet, where the planet absorbs the same amount of light from the star each orbit and does so symmetrically about the star, modulo stellar inhomogenities.
The time variability of the flux dips indicates that material with a projected area $\sim$1\% of the stellar surface is created and cleared in $\lesssim$ 1 day timescales in order to extinct $\sim$1\% of starlight at one epoch and $\sim$0\% at another.
These unusual light curve behaviors are best explained as dust extinction from a comet-like tail of material that is escaping a rocky planet (KIC 12557548 b, hereafter \sha) in a short period orbit.
The comet-like shape of the dust explains the asymmetric shape of the light curve profile.
The KIC 12557548/KOI-3794/Kepler-1520 system (hereafter \shStar) has the exciting possibility of being an opportunity to study a planet that has been peeled away layer by layer to reveal an interior core.

After the discovery of \replaced{K1255}{\shStar}, other systems were found to show similar light curve behavior: K2-22\deleted{b} \citep{sanchis-ojedak2-22}, WD 1145+017 \citep{vanderburg2015wdDisintegrating} \replaced{,}{and} KOI 2700 \citep{rappaport2014KOI2700} \deleted{and KIC 3542116 \citep{rappaport2018exocomets}}.
A collection, albeit small, of several different disintegrating planet systems makes for a useful laboratory for understanding planet evolution and composition.
Multi-wavelength light curve measurements show that the dust particles inferred to be escaping from these objects can have wavelength-dependent transmission, as predicted for sub-micron dust particles \citep{bochinski2015evolving,sanchis-ojedak2-22}.

Furthermore, there are perplexing systems that also display stochastic transit behavior but for which there are still no fully agreed-upon explanations: KIC 8462852 (ie. Boyajian's star) \citep{boyajian846}, RIK-210/EPIC 205483258 \citep{david2017rik210}  and PTFO 8-8695 \citep{vanEyken2012ptfTTauri}.
KIC 8462852 may be a family of comets from one or more parent bodies that recently broke apart \citep{boyajian846,bodman2016kic8462852cometFamily,wright2016familiesOfSolutions}.
\added{Similarly, a new system (KIC 3542116) has been discovered that has a light curve explained by extrasolar comets that produce transits in broadband optical light \citep{rappaport2018exocomets}.}
RIK-210 and PTFO 8-8695 are both in young systems ($\lesssim$ 10 and $\lesssim$3 Myr, respectively), so they may be proto-planet candidates or structures within a protoplanetary disk \citep{stauffer2017orbitingClouds,yu2015ptfO8d8695PlanetHypothesisTest}.
RIK-210 could be accreting material that causes variable extinction while PTFO 8-8695's orbit may be precessing and causing variations in transit depth depending on its longitude, though only portions of the lightcurve are explained this way \citep{barnes2013PTFO8d8695NodalPrecession,yu2015ptfO8d8695PlanetHypothesisTest}.

The \kepler\ spacecraft performed near-continuous long cadence photometry of KIC 12557548 during its entire main mission lifetime from 2009 to 2013.
During this time, the transit depths averaged around $\sim$0.6\% but varied stochastically from just one 15.7 hour orbit to the next.
\citet{vanWerkhoven2014} analyzed 15 of the total 17 quarters of data and calculated every transit depth during this period.
There were two intervals near obits 50 and 1950 that showed shallower (0.1\%) than average transit depths for roughly one month in duration.

The underlying planet candidate \sha\ has not been detected directly, but several observational constraints have placed upper limits on its size and mass.
High precision radial velocity measurements of the host star with Keck/HIRES put an upper limit on the reflex motion due to the planet and constrains the mass to be $\lesssim 1.2 M_{\mathrm Jup}$ \citep{croll2014}.
\citet{masuda2018rvKIC1255} find that the upper limit on the radial velocity semi-amplitude is 86 m/s corresponding to a planet mass $\lesssim 0.28 M_{\mathrm Jup}$ using the Subaru High Dispersion Spectrometer (HDS).
There is no detection of secondary eclipses of \sha\ with a 3$\sigma$ upper limit of $5 \times 10^{-5}$ in the \kepler\ bandpass, which means that for an albedo of 0.5, the radius of the planet must be smaller than 4600 km \citep{vanWerkhoven2014}.
A radiative-hydrodynamic wind model predicts that the mass loss rate will be a strong function of planet mass and for the present-day mass loss rate of $\sim$ M$_\oplus/Gyr$ (determined by the mass of dust needed to absorb and scatter 10$^{-2}$ of star light per orbit), the modeled mass is 0.014~$M_\oplus$ \citep{perez-becker}.

One possible explanation for the planet disintegration is that it is tied to stellar activity.
\citet{kawahara2013starspots} found an anti-correlation between stellar flux and the depth of transit events and suggested that the alignment of the planet's position (true anomaly) with active regions on the star causes disintegration.
The alignment of spots and disintegration activity could be due to XUV photoevaporation or magnetic reconnection events.
The anti-correlation between stellar flux and transit depth was confirmed by \citet{croll2015starspots}, though the authors find that occultations of star spots can also modulate the transit depth at the 22.9 day rotation period of the star.
\added{We will revisit stellar activity in Section \ref{sec:activity}.}
\deleted{Recently, \citet{rackham2018transitSourceEffect} have studied the effect of un-occulted spots on the transit depths exoplanets of M stars, the ``transit source light effect.''
For the observed amplitude variations of 0.8\% \citep{kawahara2013starspots}, scaling relations for K4 V dwarfs indicate that transit depths of exoplanets can be biased by 10$^{+20}_{-5}$\% depending on the distribution and sizes of spots if the transit chord is completely devoid of spots.
The large physical size of \sha\, as indicated by a transit duration longer than the time to cross the star, means that spots are likely to be crossed and diminish the transit source effect, so \sha's 30\% transit depth variation would require a worst-case starspot coverage to be explained by un-occulted spots.}

The \kepler\ data was used to put constraints on the particle size distribution and composition of the dust particles.
\citet{budaj12} and \citet{brogi2012} modeled the light curve, which includes a slight increase in flux before the transit begins.
This pre-ingress flux increase is caused by forward scattering dust particles and the scattering function is sensitive to particle size modulo composition.
\citet{budaj12} and \citet{brogi2012} find particle sizes $\sim$0.1 to $\sim$1.0$\mu$m in size based on this method.
The tail length can also be used to constrain the composition of the particles and \citet{vanlieshout2014kic1255comp} find that the sublimation times of corundum and iron-rich silicates are consistent with the observed tail lengths.

Multiple wavelength monitoring can also constrain the dust particle sizes because the scattering by dust is a strong function of wavelength.
\citet{bochinski2015evolving} observed wavelength dependence in the transits by \sha\ and find that the extinction is similar to the ISM particles with sizes from 0.25 to 1~$\mu$m.
Subsequently, \citet{croll2014} obtained a $K'$-band light curve simultaneously with the \kepler\ spacecraft to put a constraint on the dust extinction as a function of wavelength.
\citet{croll2014} find a particle size distribution of $\gtrsim 0.5\mu$m.
Recent modeling by \citet{ridden-harper2018chromaticKIC1255} shows that the wavelength-dependent extinction function could be due to the tail's changing optical thickness.

\citet{schlawin2016kic1255} observed 8 total transits of the system for 4 nights in 2013 and 4 nights in 2014 for the $r'$ band using the MORIS imager \citep{Gulbis2011} on SpeX/IRTF \citep{rayner03}.
The SpeX/IRTF spectrum, though affected by systematic noise, confirmed the \citet{croll2014} result that particle sizes are $\gtrsim 0.5$~$\mu$m for a pyroxene composition.
\citet{schlawin2016kic1255} found that the optical transit depths were all weaker than $0.43\%$ and that the probability of this randomly occurring based on random behavior from \kepler\ statistics was around 0.2\%.
This was evidence that the observations occurred during weaker periods (as found in \citet{vanWerkhoven2014}'s 15 \kepler\ quarter analysis) or that the disintegration activity was falling off with time.
We performed follow-up $R$ band photometry of the \shStar\ system to explore these two hypotheses and see if the the disintegration has returned to the levels found in the \kepler\ mission and discuss this in Section \ref{sec:photObs}.
We also re-analyzed the \kepler\ photometry in order to quantify the light curve behavior in comparison to ground-based photometry in Section \ref{sec:Kepler}.

The high resolution spectrum of \shStar\ \citep{kawahara2013starspots} showed a different temperature and gravity than photometric methods \cite[e.g.][]{brown2011kic,huber2014kicprop}.
\citet{vanlieshout2016kic1255} suggest that one exciting possibility is that gases from the disintegrating planet contaminate the high resolution spectrum.
This would enable compositional and kinematic studies of the disintegrating planet.
In Section \ref{sec:stellarCharacterization}, we examine archival Subaru High Dispersion Spectrograph (HDS) spectra of the star to assess the stellar parameters and explore the possibility that sublimated gases appear in the high resolution spectrum.
\added{In Section \ref{sec:activity}, we examine how un-occulted starspots can affect the transit depth behavior.}
We conclude in Section \ref{sec:conclusions}.

\begin{deluxetable*}{lrrrr}
\tablecaption{List of photometric observations}\label{tab:photSummary}
\tablewidth{0pt}
\tablehead{
\colhead{Date} &
\colhead{Weather} &
\colhead{Airmass} &
\colhead{Seeing} &
\colhead{Amplitude, $A$} \\
\colhead{UT} & & & (\arcsec) & \\
 }
\startdata
2016 Jun 10 & Cloudy	& \nodata & \nodata &  \nodata \\
2016 Jun 12 & Clear 	& 1.29 $\rightarrow$ 1.06 $\rightarrow$ 1.10	& 1.3 &	0.94 $\pm$ 0.16 \\
2016 Jun 14 & Clear 	& 1.47 $\rightarrow$ 1.06 $\rightarrow$ 1.12	& 1.4 &	0.70 $\pm$ 0.18\\
2016 Jul 12 & Clear 		& 1.19 $\rightarrow$ 1.06 $\rightarrow$ 1.63	& 1.2 &	0.78 $\pm$ 0.14 \\
2016 Jul 14 & Mostly clear	& 1.24 $\rightarrow$ 1.06 $\rightarrow$ 1.55 & 1.2 & 	1.33 $\pm$ 0.14 \\
2016 Jul 16 & Clear 		& 1.36 $\rightarrow$ 1.06 $\rightarrow$ 1.26	& 1.4 & 	2.14 $\pm$ 0.19 \\
\enddata
\tablenotetext{}{Summary of the attempted photometric observations of \shStar\ with the 61-inch Kuiper telescope.
The seeing is expressed as the median FWHM for the night in arcseconds.}
\end{deluxetable*}

\section{Photometric Observations}\label{sec:photObs}
We attempted to observe 6 transit events of \sha\ with the Mont4k imager on the 61-inch Kuiper telescope located on Mt Bigelow, Arizona in 2016 using the f/13.5 Cassegrain focus.
On all nights we used 90 second exposures using a Harris-R filter to best approximate the $r'$ filter used in \citet{schlawin2016kic1255}.
5 of the 6 nights were clear with the exception of a few passing clouds on UT 2016 Jul 14.
Cloudy skies on UT 2016 Jun 10 prevented high precision time series.
Table \ref{tab:photSummary} shows a summary of our observations and the UT dates.
The data was taken in a 3x3 binning mode resulting in an effective 0.43"/pixel plate scale.

\subsection{Photometric Pipeline}

For the photometric reduction, we used the Python \texttt{ccdproc} package \citep{craig2015ccdproc} to generate master flat field and bias files.
They were combined with an average with a low threshold of 2$\sigma$ and a high threshold of 5$\sigma$.

\begin{figure*}
\begin{centering}
\includegraphics[width=0.49\textwidth]{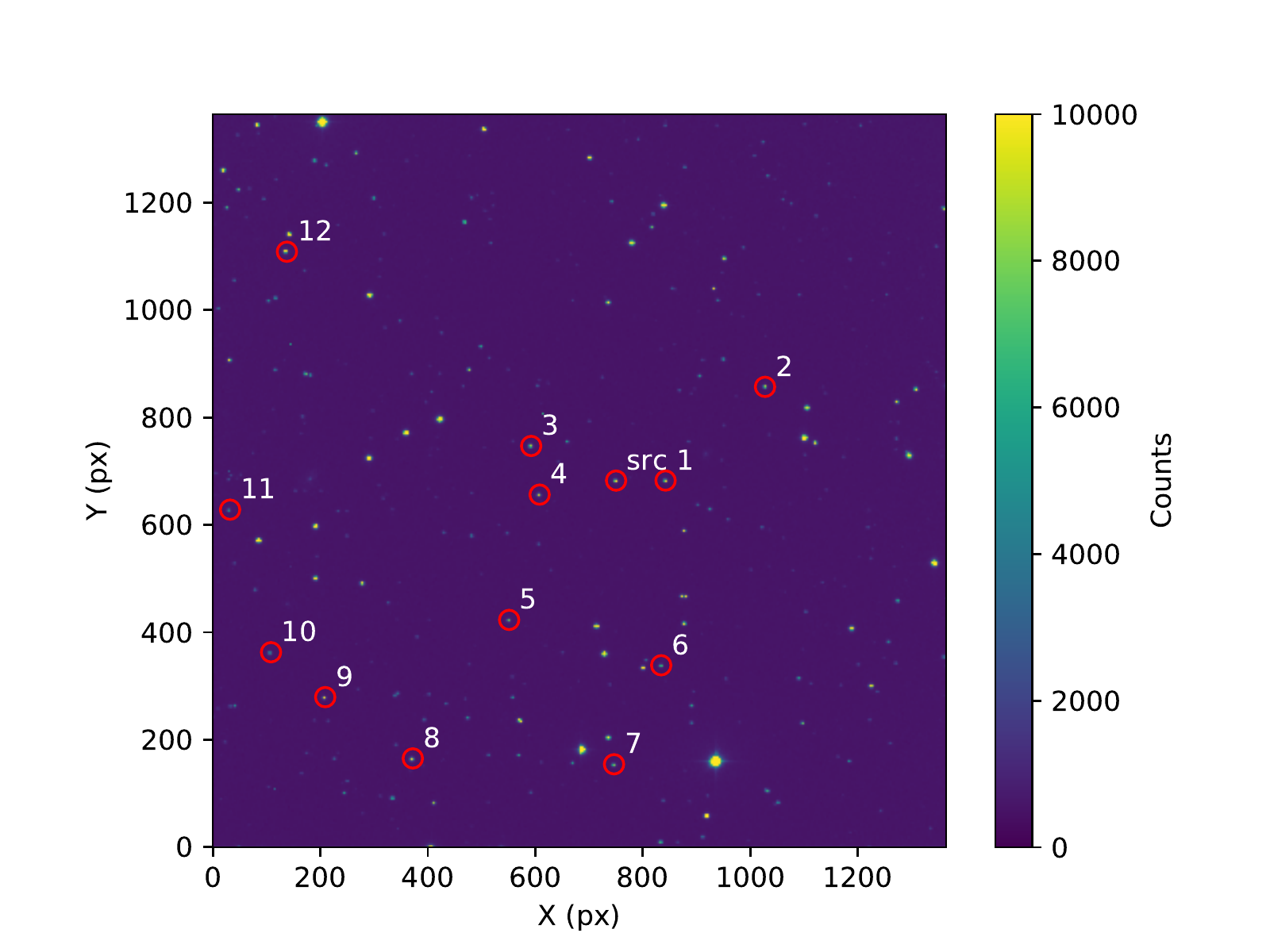}
\includegraphics[width=0.49\textwidth]{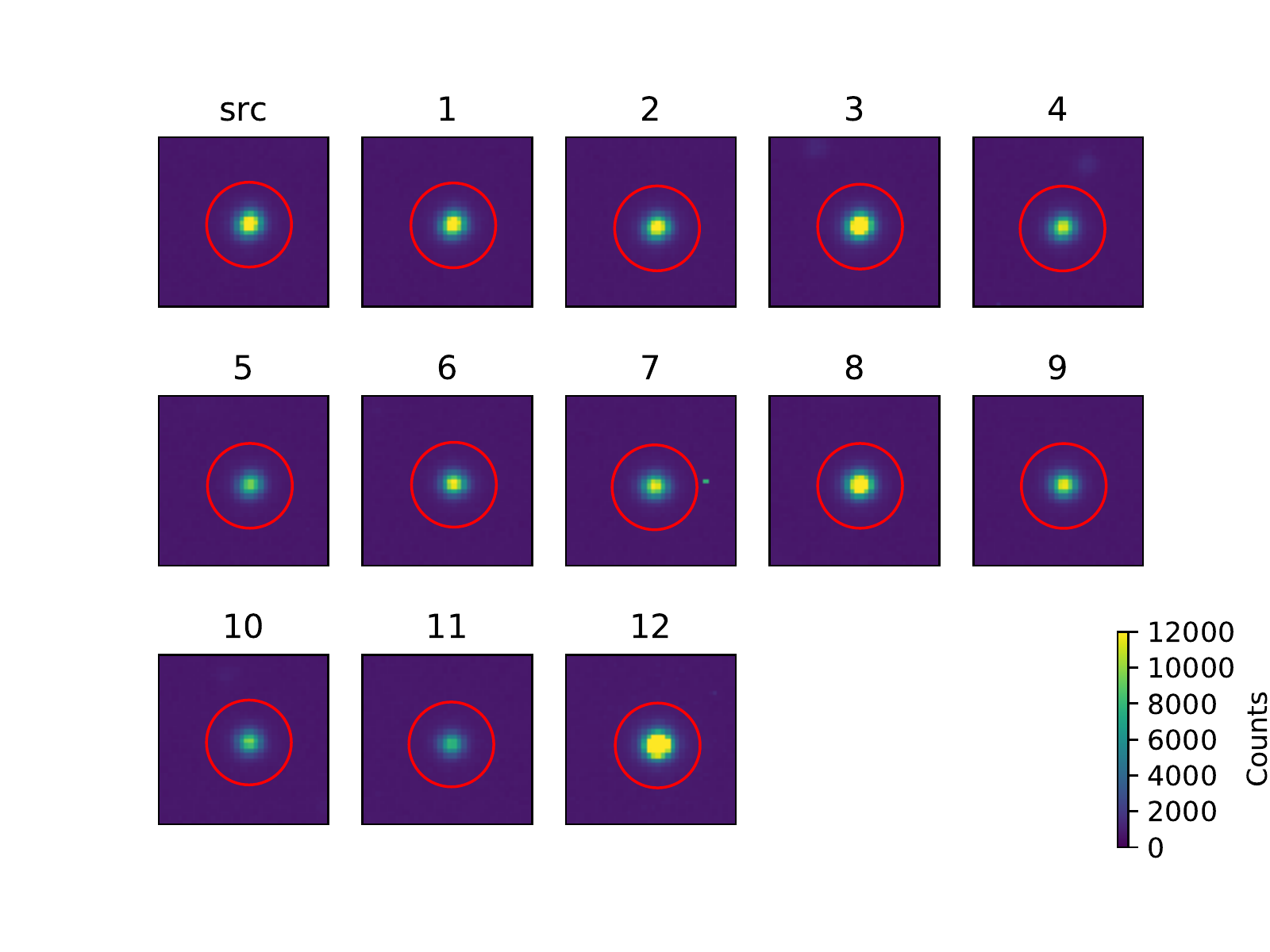}
\caption{{\it Left:} \shStar\ (src) and reference stars (numbered) used on UT 2016-07-14 are shown over the full 9.8\arcmin $\times$ 9.8\arcmin\ Field of View of \added{the} Mont4K \added{imager}.
{\it Right:} Postage stamp zoom-ins for the source and reference stars show the point spread functions for each star.
The reference stars were chosen to have similar count levels as \shStar.
}\label{fig:refStars}
\end{centering}
\end{figure*}

We used the Python \texttt{photutils v0.3} \citep{bradley2016photutilsv0p3} package to centroid and extract fluxes of the \shStar\ and the reference stars.
We used a photometric aperture of 9 pixels (3.9") and a background annulus from 9 to 12 pixels (3.9" to 5.2") on all stars and nights.
12 stable reference stars with similar R band brightnesses were used to construct a reference time series to correct for variable telluric and instrument throughput.
The reference stars are identified Figure \ref{fig:refStars} across the 9.7\arcmin $\times$ 9.7 \arcmin\ Mont4K field of view.
These 12 reference stars' time series were combined with a weighted average and this combined time series was divided into the target, \shStar.

\begin{figure*}
\begin{centering}
\includegraphics[width=0.8\textwidth]{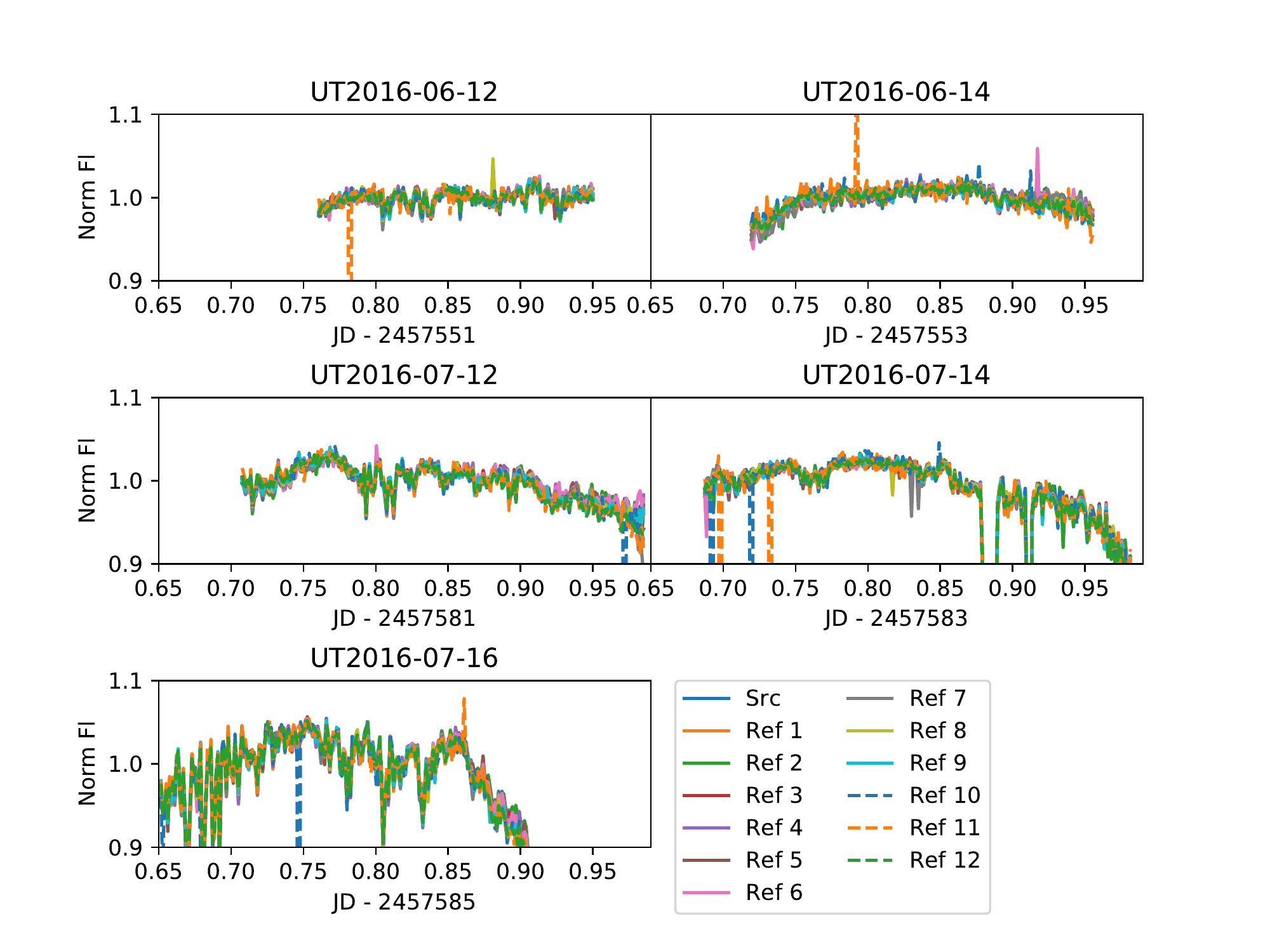}
\caption{Time series photometry for all stars and all nights.
The nights in July appear more strongly affected by cloud/and or seeing variations, whereas the two nights in June are stable to within a few percent.
All source radius, background start and background annulus end parameters were 9, 9 and 12 pixels respectively.}\label{fig:allNightallStar}
\end{centering}
\end{figure*}

All light curves for \shStar\ and the 12 reference stars are shown in Figure \ref{fig:allNightallStar}.
These are the same reference stars shown in Figure \ref{fig:refStars}.
The two nights 2016 Jun 12 and 2016 Jun 14 show greater overall stability than the 3 nights in July, 2016.
We suspect that the higher moisture levels and occasional cloud passages (such as the ones that caused huge drops in flux on UT 2016 Jul 14) affected the later observations.

\subsection{Transit Depth Fitting}\label{sec:transFitting}

\begin{figure*}
\begin{centering}
\includegraphics[width=0.8\textwidth]{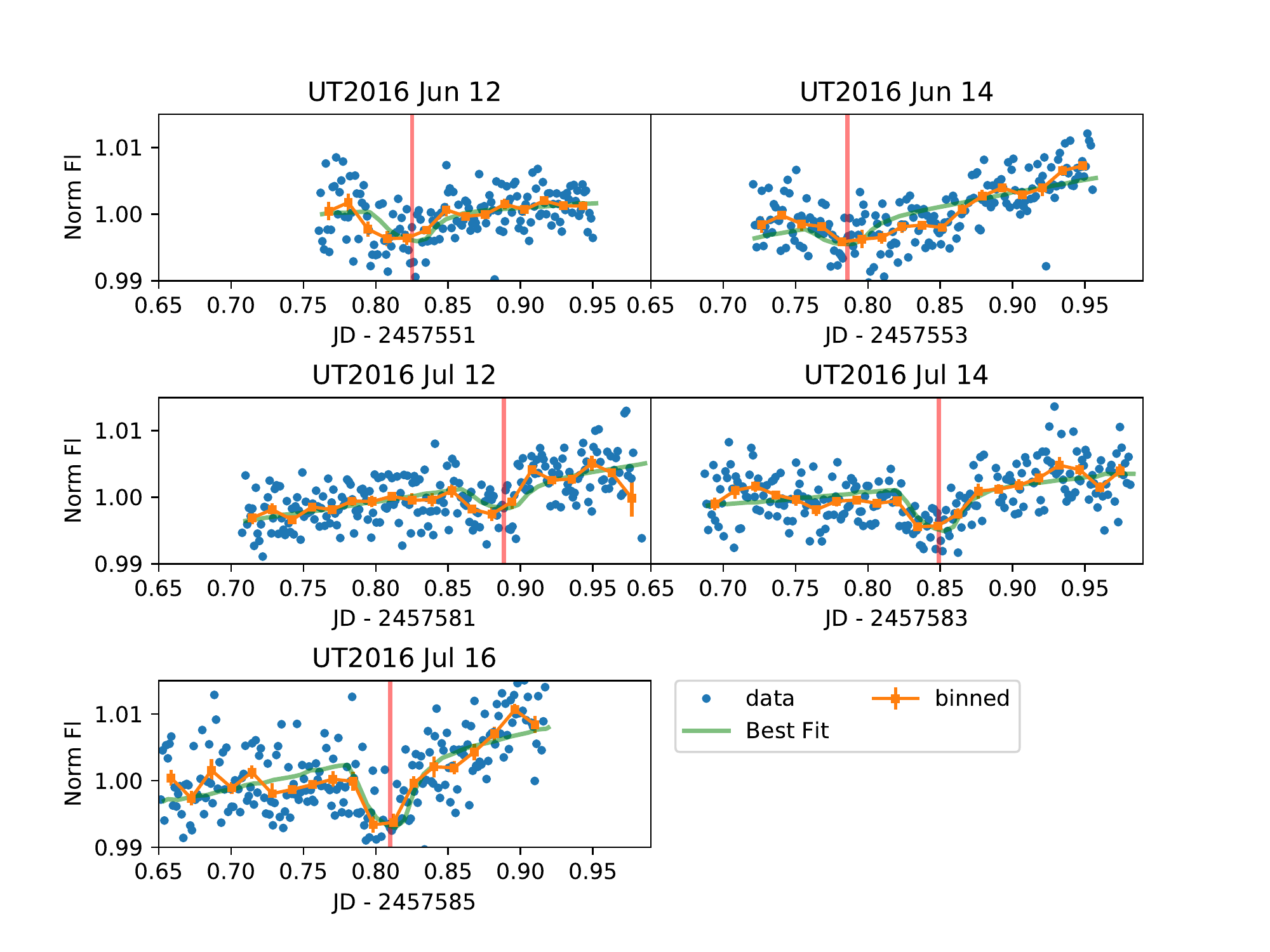}
\caption{Reference-corrected time series photometry of \shStar\ for all nights (blue circles).
Orange squares mark 20 minute flux bins for visual clarity.
The expected transit epochs from the \kepler-based ephemeris \citep{croll2015starspots} are shown as vertical red bars.
We also fit each light curve with a model that scales the average \kepler\ short cadence data (Equation \ref{eq:transFitting}) shown in green.
Transits are clearly visible for all nights except for UT 2016 Jun 14.}\label{fig:allNightrefCorrect}
\end{centering}
\end{figure*}

Figure \ref{fig:allNightrefCorrect} shows the reference-corrected time series for \shStar\ for all nights.
We use the planet transit ephemeris from \citet{croll2015starspots} where the \kepler\ flux minimum is the ``transit center'':
\begin{equation}\label{eq:ephem}
T = T_0 + n P
\end{equation}
where $T_0 = 2 454 968.982 0 \pm 0.000 7$ BJD$_{TDB}$ and $P = 0.653 553 4 \pm 0.000 000 2$ days.
The expected transit epochs from this ephemeris are shown for reference as vertical red bars.
Typical uncertainties for the \kepler\ ephemeris are $\sim$ 1 minute at these dates due to the large number of transits measured by the \kepler\ observatory over nearly 4 years.
The ``transit duration'' from \kepler\ light curves is about 0.06 days\footnote{We define the transit duration as the point at which the average light curve measured by the \kepler\ spacecraft drops below 99.95\% of the normalized flux as in \citet{schlawin2016kic1255}}, which is consistent with these transit durations.

We find the best fit to each light curve using the average \kepler\ short cadence light curve as a model.
We include a linear baseline so that the model is
\begin{equation}\label{eq:transFitting}
F(t) = \left(A \cdot f_{SC}(t) + 1 \right)  (1 + B + C \cdot t),
\end{equation}
where $F(t)$ is the normalized flux at orbital phase $t$, $A$ is the amplitude parameter, $f_{SC}(t)$ is average \kepler\ short cadence light curve linearly interpolated at an orbital phase $t$, $B$ is the baseline offset and $C$ is the baseline slope.
We construct $f_{SC}(t)$ by phasing the \kepler\ time series with Equation \ref{eq:ephem}, dividing by a quadratic baseline from each transit, averaging the results and subtracting 1.0 to get the differential flux.
We use the Markov Chain Monte Carlo (MCMC) package \texttt{emcee} \citep{foreman-mackey2013emcee} to find the transit depth amplitude and uncertainty.
We use 50 MCMC chains with 500 burn-in points and 1500 points to sample the posterior distribution.
The best-fit light curves are shown in Figure \ref{fig:allNightrefCorrect} and the best-fit amplitudes are listed in Table \ref{tab:photSummary}.

To ease the visual comparison of the models and measured data, we bin the 61-inch Mont4K data into 20 minute long bins, shown in Figure \ref{fig:allNightrefCorrect}.
Obvious cosmic ray outliers were discarded by removing points (at any epoch within the reference-corrected time series) with fluxes below 0.98 and above 1.02 times the median flux for the night.
The error bars for the binned data are calculated from the standard error in the mean for each bin.
When calculating the best fit model and posterior distribution of parameter, however, the un-binned data is used.

In contrast to \citet{schlawin2016kic1255}, the transit depths in 2016 June and 2016 July are similar to the distribution measured by the \kepler\ spacecraft from 2009 to 2013.
\citet{schlawin2016kic1255} found significantly weaker transit depths that the \kepler\ spacecraft in 2013 Aug-Sep and 2014 Aug-Sep, which indicated that either disintegration activity was slowing down or that there were quiescent periods during those observations.
The transit depths listed in Table \ref{tab:photSummary} are ``back to normal'' in that they are consistent with the \kepler\ results.
We will revisit the statistics of these events in Section \ref{sec:statistics}.

\section{\kepler\ Re-analysis}\label{sec:Kepler}

We examine the 17 total quarters of \kepler\ data to understand the statistics of the transits and put the 2013-2016 observations in context. Mainly, we sought to evaluate the likelihood of the weak transits observed in 2013-2014 and normal transits in 2016.
We downloaded all 17 quarters of \kepler\ Long-Cadence data from the MAST archive beginning on 2009 May 13 and ending on 2013 April 08 for a total of 1425 days or 2181.5 orbits.
The \kepler\ data were examined for one transit or secondary eclipse at a time with a window of $\pm$0.28 in orbital phase.
A quadratic baseline is fit to all the points before a phase of -0.12 and after a phase of 0.15 within the $\pm$ 0.28 phase window.
The -0.12 and 0.28 phases are set from the points within the average \kepler\ light curve that are within $\pm$ 200 ppm from the median out-of-transit flux.
All points are divided by this baseline to produce a normalized flux transit or eclipse profile.
The secondary eclipses are used to quantify the noise in the transit depths.

\subsection{\kepler\ Secondary Eclipse}

\begin{figure*}[!hbtp]
\begin{centering}
\includegraphics[width=0.45\textwidth]{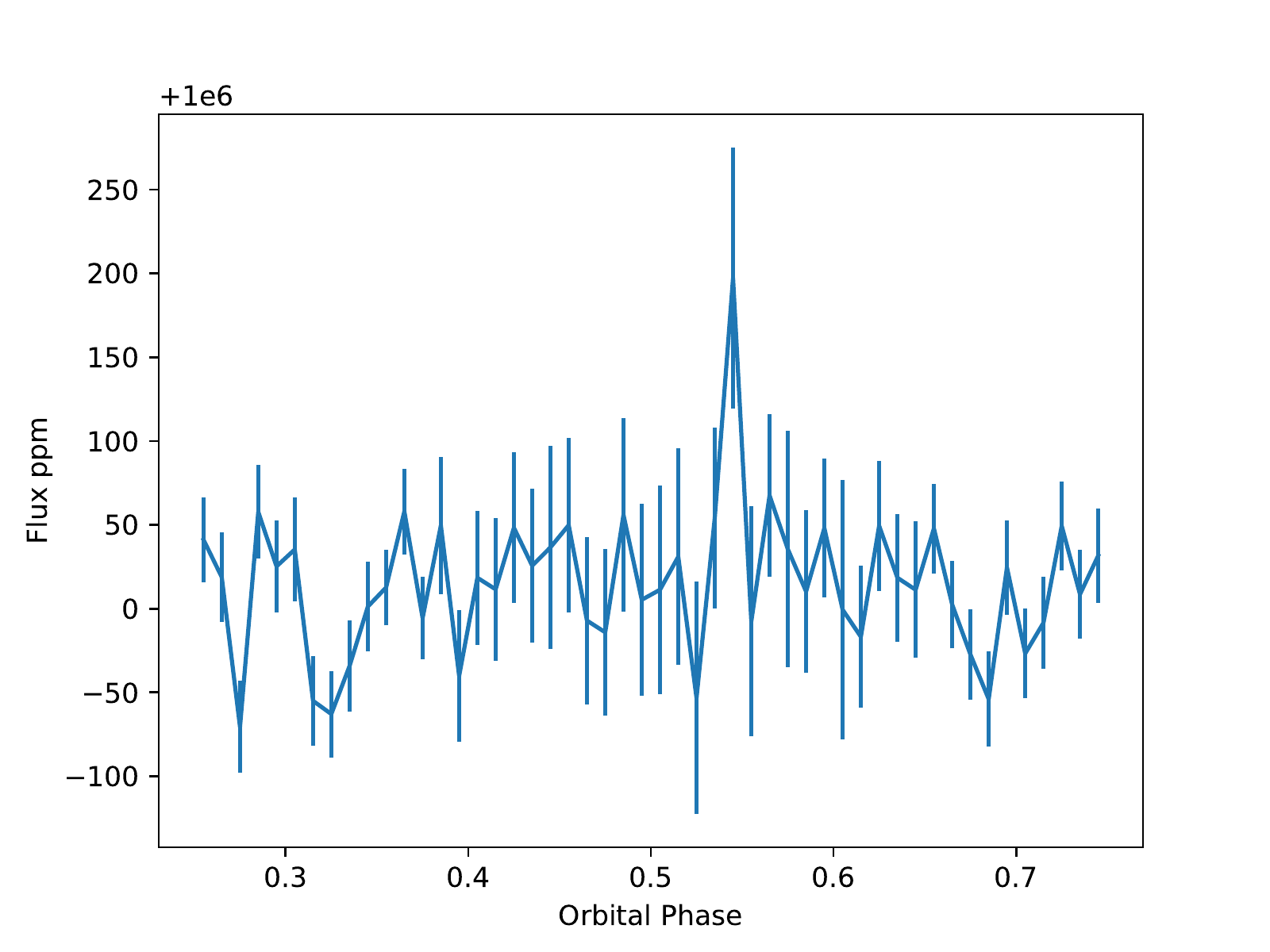}
\caption{No secondary eclipse (f$_p$/f$_*$ $\lesssim$ 50 ppm) is detected from a phase range of 0.25 to 0.75, even with 17 quarters (4 years) of \kepler\ Long Cadence data.
Here, the data are binned to orbital phases of 0.016 (15 minutes) and errors calculated from the standard error in the mean of the data within a bin.
}\label{fig:secEclipse}
\end{centering}
\end{figure*}

We confirm the result of \citet{vanWerkhoven2014} that no secondary eclipse is detected at f$_p$/f$_*$ $\lesssim 50$~ppm, where f$_p$/f$_*$ is the relative flux of the planet to star in the Kepler bandpass, as seen in Figure \ref{fig:secEclipse}.
We then use the secondary eclipses as a way to characterize the noise of the primary transits.
Both have the same quadratic baseline removal process over the same time intervals.

\begin{figure*}[!hbtp]
\begin{centering}
\includegraphics[width=0.32\textwidth]{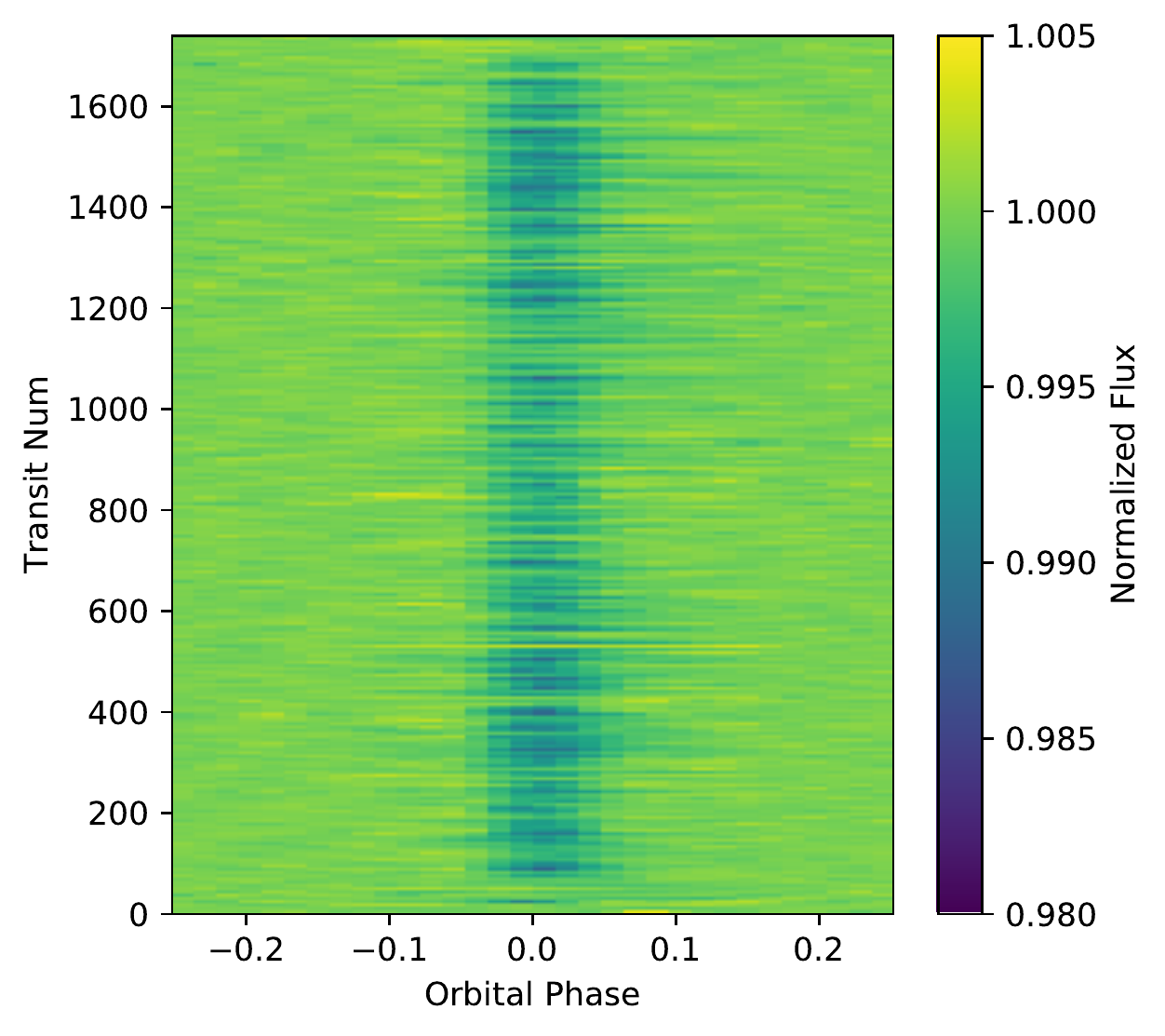}
\includegraphics[width=0.32\textwidth]{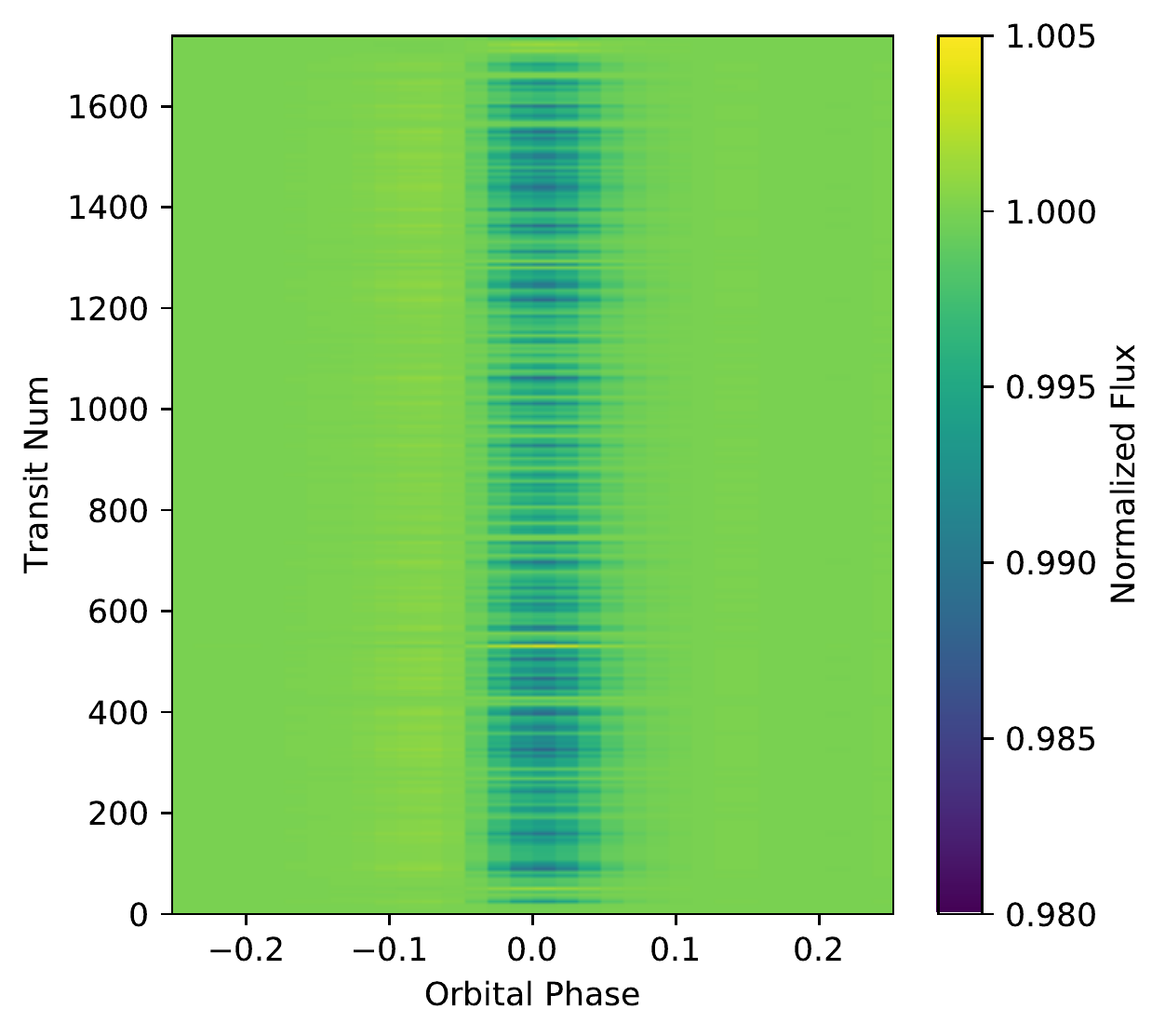}
\includegraphics[width=0.32\textwidth]{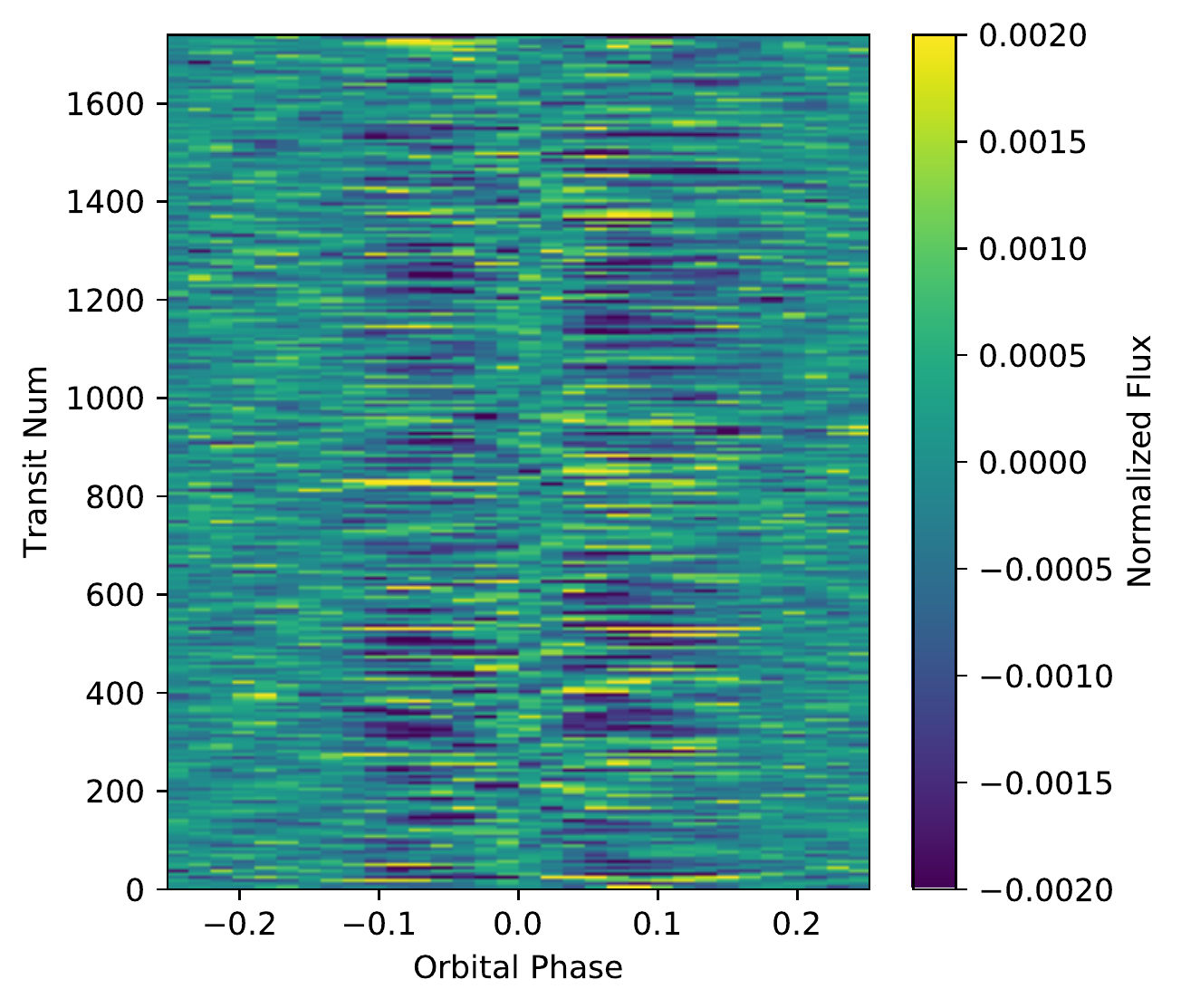}
\caption{Cleaned \kepler\ Long-cadence data (Left) and model fit using the average light curve (Middle).
The residuals (Right) show significant deviations near phases of -0.05 and 0.05.
}\label{fig:riverPlots}
\end{centering}
\end{figure*}

\subsection{Average Light Curve Analysis}

In order to perform some analysis techniques on the data and to study the transit depth behavior, we first create a uniform cleaned two dimensional array of the time series.
We linearly interpolated the \kepler\ long-cadence data set onto a fixed phase grid with a spacing of 15 minutes (to Nyquist sample the 30 minute observation cadence) and put together all the time series that have no gaps into this two dimensional grid.
This cleaned \kepler\ long-cadence array is shown in Figure \ref{fig:riverPlots}.

We start by taking the average transit light curve of all the long cadence data (the mean of the 2D grid along the transit number axis).
We then take the dot product ($\bullet$) with this average light curve to create an amplitude time series.
\begin{equation}
A' \equiv \frac{(f - 1) \bullet \vec{a}}{\vec{a} \bullet \vec{a}},
\end{equation}
where $\vec{a}$ is the amplitude vector, $f$ is the flux matrix with x indices of time and y indices of transit number and $a$ is the average spectrum.
This dot product, $A'$ is comparable to the transit amplitude $A$ in Equation \ref{eq:transFitting}.

This dot product is then multiplied by the average transit light curve to create a 2D model grid.
The residual of this model (Figure \ref{fig:riverPlots}) shows significant $\pm$0.2\% deviations from white noise, indicating that the dusty tail surrounding \sha\ changes in shape (or scattering properties) with time.
We explore more sophisticated models than the average light curve below that are based on principal component analysis.
However, this average model serves as a useful comparison to fitting the transit depth for ground-based photometry, as performed in Section \ref{sec:transFitting}.
The 0.2\% residuals with this average model are comparable with the typical photometric scatter of $\sim$ 0.1\% to 0.2\% for 20 minute timescales in \citet{schlawin2016kic1255} and this work.
We statistically compare the transit depths from the \kepler\ mission from 2009 to 2013 to the ground-based measurements from 2013 through 2016 in Section \ref{sec:statistics}.

\subsection{Principal Component Analysis}
\begin{figure*}[!hbtp]
\begin{centering}
\includegraphics[width=0.32\textwidth]{photometry_riverplot.pdf}
\includegraphics[width=0.32\textwidth]{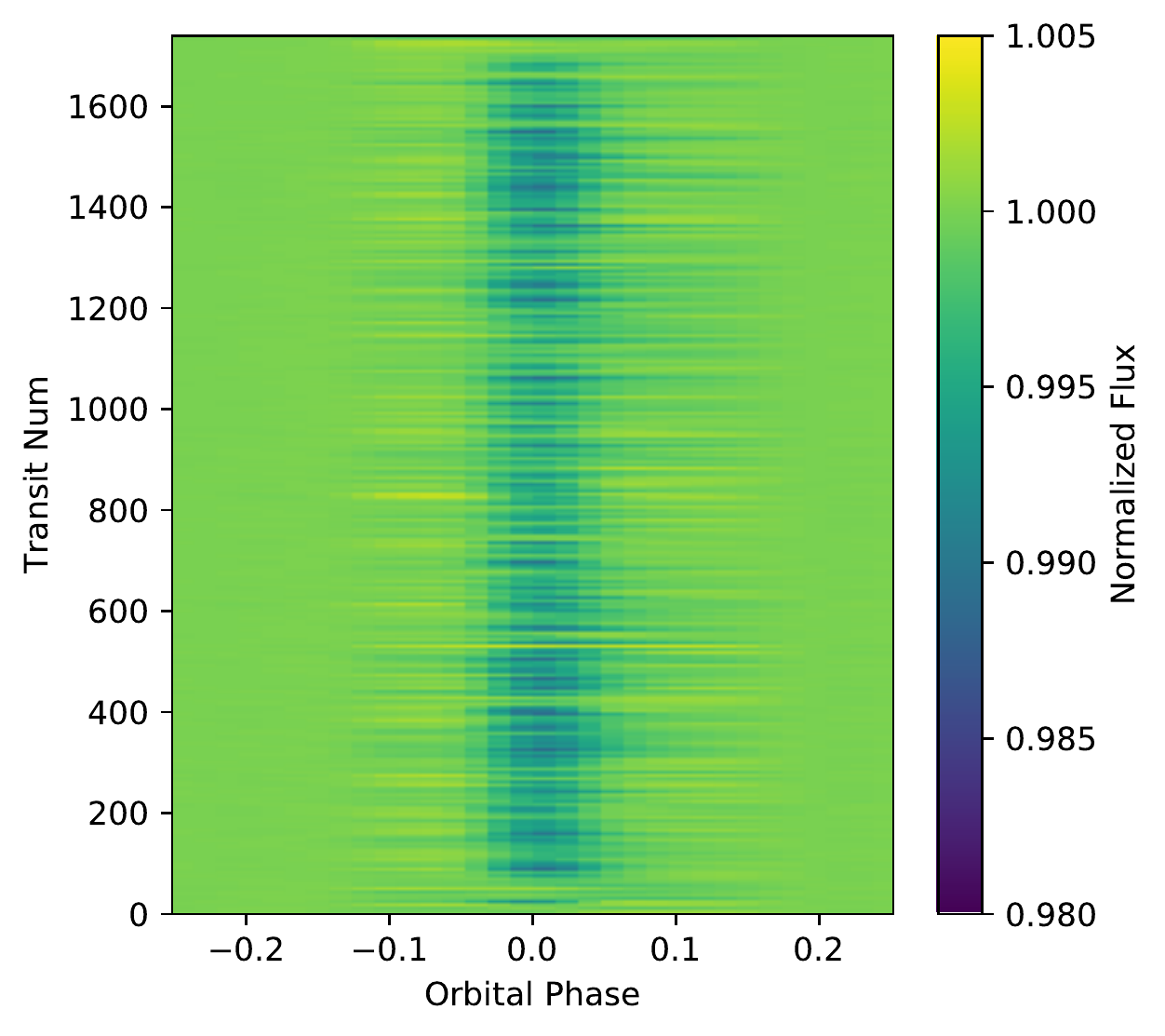}
\includegraphics[width=0.32\textwidth]{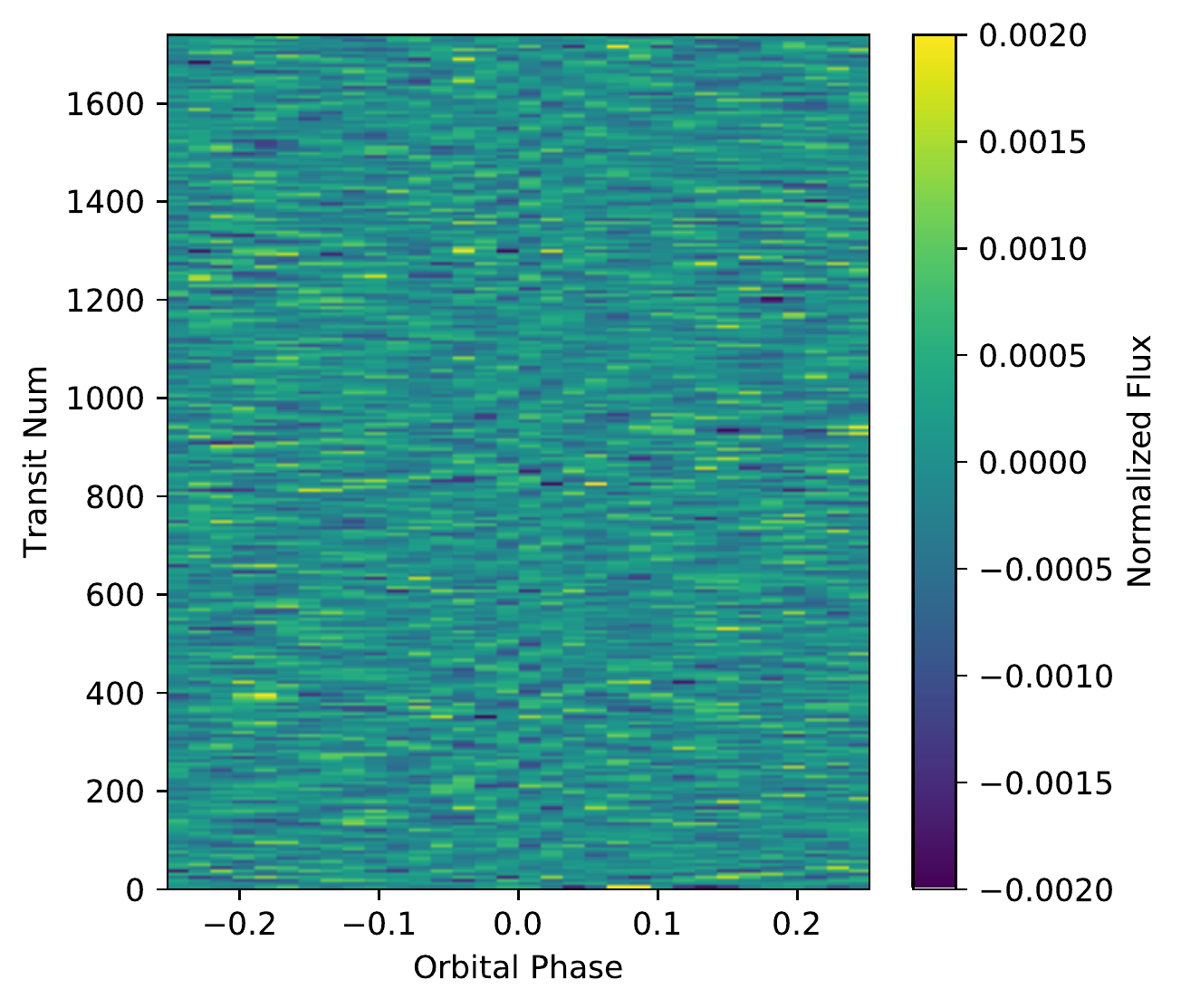}
\caption{Cleaned \kepler\ Long-cadence data (Left) and model fit using 4 Principal Components (Middle).
The residuals are $\lesssim 0.1\%$ (Right), significantly smaller than the $\sim 0.2\%$ residuals when using the average-light curve method shown in Figure \ref{fig:riverPlots}.
}\label{fig:riverPlotsPCA}
\end{centering}
\end{figure*}

As another way of analyzing the light curve data, we apply Principal Component Analysis (PCA) to this two dimensional grid.
Here, we are assuming that the grid of orbital phases (spaced by 15 minutes) contains random correlated variables with different realizations along the transit number axis.
The first principal component eigenvector is the linear combination of these random variables that maximizes the variance in the flux.
The second principal component eigenvector is the linear combination of these random variables with the next-highest variance while being orthogonal to the first principal component.
This continues up to some finite number of principal components, with the hope that we can use a few orthogonal light curves to adequately describe the data \citep[e.g.][]{jolliffe2002pca}.
We use the Python package \texttt{scikit-learn} \citep{pedregosa2011scikit-learn} to calculate the principal components and eigenvectors.
\begin{figure*}[!hbtp]
\begin{centering}
\includegraphics[width=0.49\textwidth]{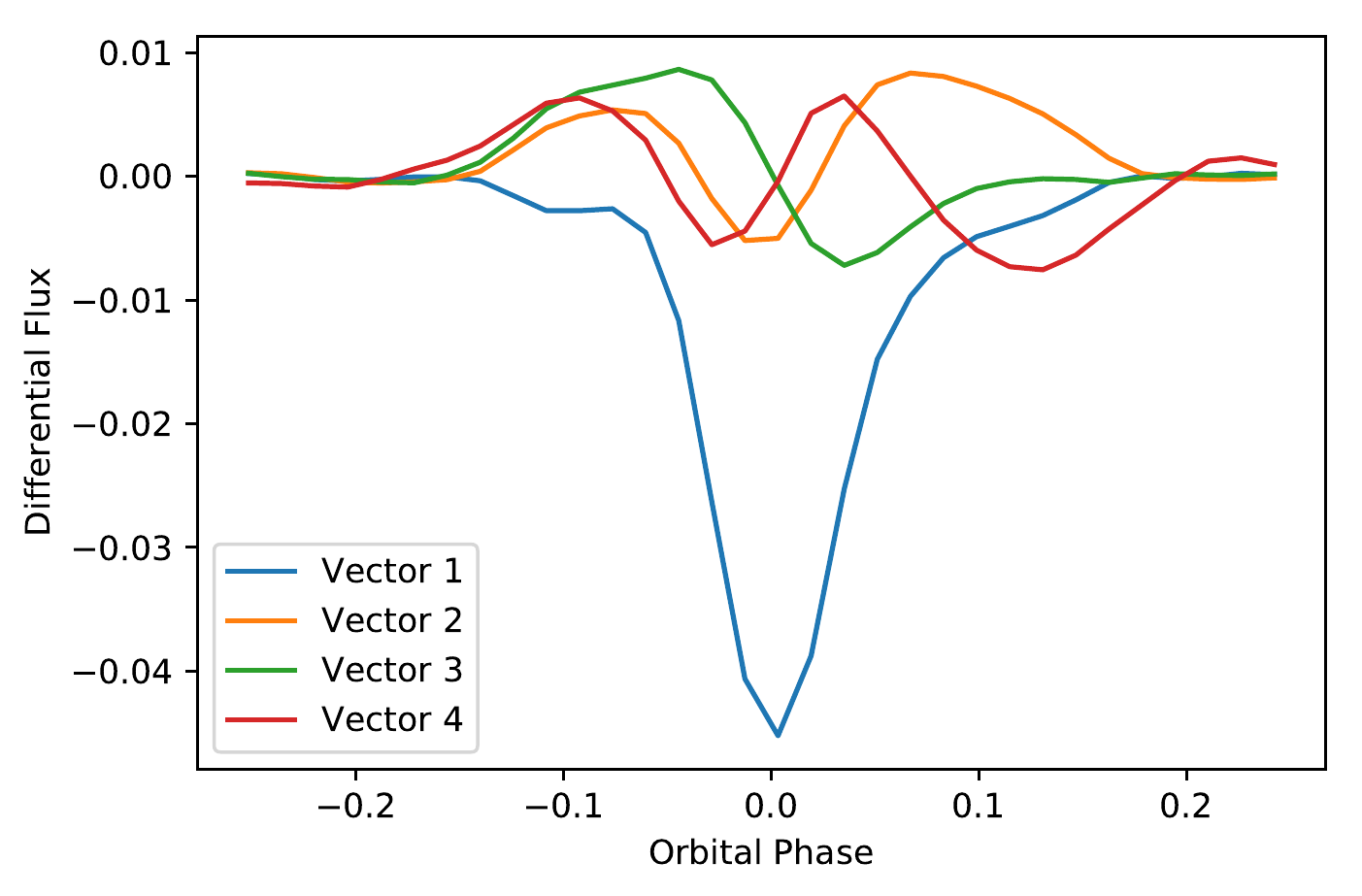}
\includegraphics[width=0.49\textwidth]{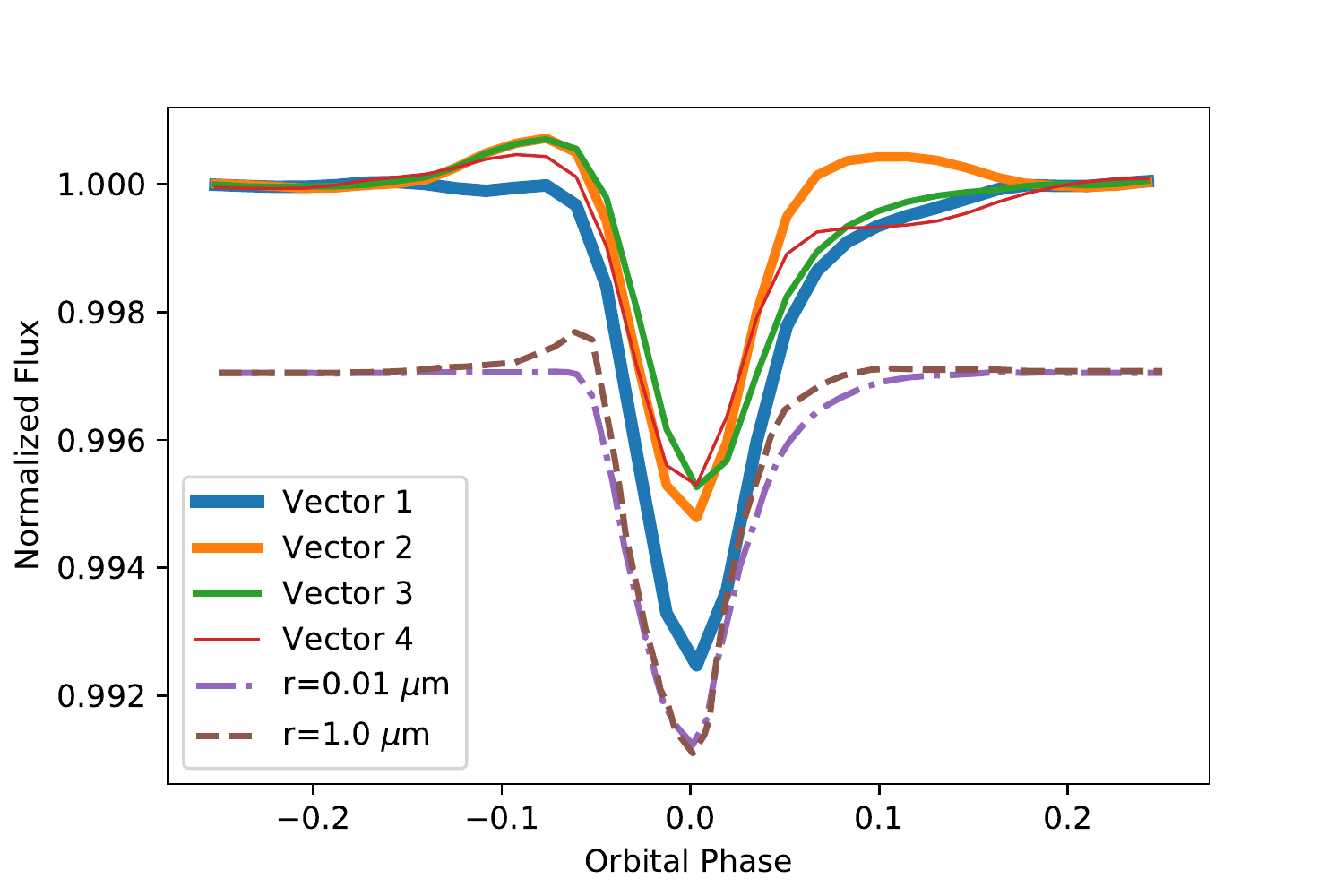}
\caption{The principal component eigenvectors scaled by their eigenvalues (Left), shows that the first eigenvector largely describes the extinction while the second shows the contributions from forward scattering.
When the eigenvectors are transformed into light curves (Right), it can be seen that the first eigenvector corresponds to small dust particle sizes with negligible forward scattering \citep[dash-dot purple model][]{budaj12}.
The second eigenvector resembles models with large dust particle sizes with pronounced forward scattering before and after the transit mid-point \citep[dashed brown model][]{budaj12}.
The models are offset vertically for clarity.
}\label{fig:pcaVectorsComp}
\end{centering}
\end{figure*}

We use a covariance matrix as opposed to correlation matrix to calculate the principal component eigenvectors.
In other uses of PCA, the variables are often scaled so that they all have a variance of 1.0 to ensure that variables with intrinsically larger variance (such as height in inches over weight in kilograms) do not get higher loadings in the principal component eigenvectors \citep{jolliffe2002pca}.
However, we do not scale the variables (columns) by variance and use a covariance matrix because all of our variables are the same units (normalized flux).
If we scaled each column in this 2D grid by its standard deviation or used a correlation matrix, this would magnify the contribution of photon noise to out-of-transit data over the real astrophysical variability of the inner phases (between $\sim$ -0.12 to $\sim$0.15).

\added{We model the 2D {\it Kepler} flux grid shown in Figure \ref{fig:riverPlotsPCA} (Left) with a set of principal components and examine the residuals to evaluate the number of principal components to keep.
We begin with 1 principal component and progressively added one at time until the residuals along the transit number axis of Figure \ref{fig:riverPlotsPCA} are the same as the out-of-transit flux.
We find that after 4 eigenvectors, the in-transit (with phases between -0.12 and 0.15) matches the out-of-transit standard deviation of $\sim 0.05\%$, also found in the secondary eclipse 2D flux grid.}
The model using the first 4 principal components shown in Figure \ref{fig:riverPlotsPCA} \added{(Right)} better matches the data and has residuals much closer to white noise than the model that uses the dot product with the average light curve.

Figure \ref{fig:pcaVectorsComp} shows the first 4 eigenvectors as a function of orbital phase.
For the eigenvectors plot, we multiply each eigenvector by its eigenvalue to emphasize the relative covariances between the principal components and phase grid variables.
The first 4 eigenvectors explain 62\%, 12\%, 6\% and 3\% of the variance (a total of 84\%) for the 32 points (variables) in the phase grid from -0.25 to 0.25.
We also transform the eigenvectors into light curves by multiplying each eigenvector by the standard deviation of principal components and adding it to the average light curve, as shown in the right side of Figure \ref{fig:pcaVectorsComp}.

The first eigenvector largely corresponds to models of small dust grains with more isotropic scattering so that the light curve is dominated by extinction \citep{budaj12,brogi2012}.
We also note that the second eigenvector resembles models of large dust grains of pyroxene from \citet{budaj12}, where the particles that are a similar order of magnitude as the wavelength exhibit forward scattering, which causes a pronounced pre-ingress flux increase.
The second eigenvector has a broader forward scattering peak than the \citet{budaj12} models, so it may be the result of a broader distribution of particles.
The fact that the pre-ingress flux increase is not found in the first eigenvector and instead is pronounced in orthogonal eigenvectors suggests that the dust particle sizes may be changing in time.
We would expect that a dust tail with a variable number of \deleted{of} particles but a constant grain size would \deleted{be} exhibit a forward scattering in the first principal component because the forward scattering flux would scale proportionally with the number of particles and transit depth.
The implication of the PCA then is that the dust particle sizes may be changing with time.
\added{We assess the uncertainty of this conclusion through the time series of the principal components.}

\begin{figure*}[!hbtp]
\begin{centering}
\includegraphics[width=0.99\textwidth]{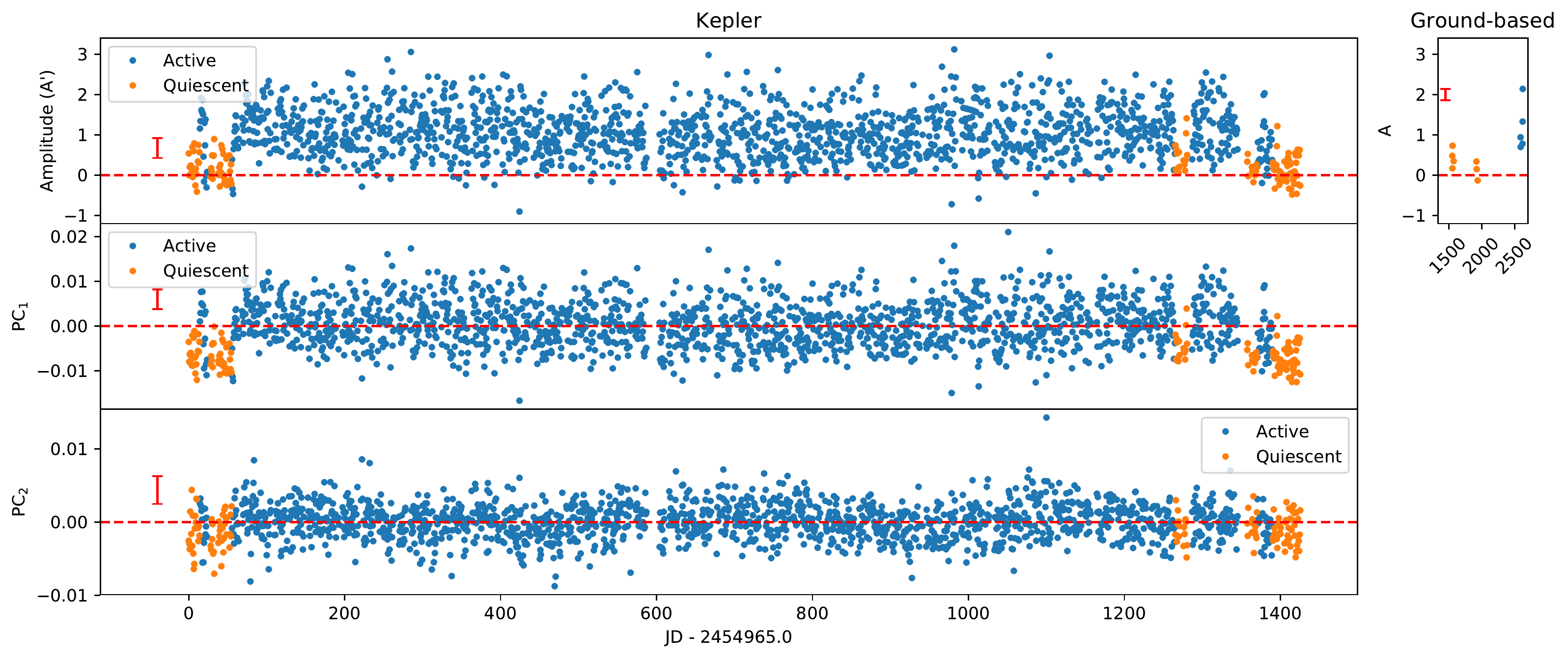}
\caption{Time series of the transit amplitude and principals components of the \replaced{Kepler}{{\it Kepler}} data (Left panel).
We also show the ground-based photometry from \citet{schlawin2016kic1255} and this work (Right upper plot).
Quiescent intervals with low transit activity are identified from the amplitude \added{and highlighted in orange. 
Typical error bars for the time series are shown as red error bars at arbitrary positions.}
The first principal component tracks closely with the amplitude.
The second principal component shows periodicity at 491 days, indicating possible long-term evolution of dust grain particle sizes.
}\label{fig:tserPC}
\end{centering}
\end{figure*}

Figure \ref{fig:tserPC} shows the time series of the first two principal components (PC$_1$ and PC$_2$).
We highlight the quiescent periods, which are also discussed in \citet{kawahara2013starspots}, \citet{vanWerkhoven2014} and \citet{croll2015starspots}.
\added{We estimate the errors in the principal components by measuring the dot product with the secondary eclipse light curve and calculating its standard deviation.}
The first principal component largely tracks closely with the dot product of the average light curve ($A'$).
The quiescent periods show up as 14 to 36 day long intervals of low amplitude events.
The second principal component shows some sinusoidal behavior at a $\sim$500 day period, indicating that the large particles may be evolving.
However, the timescale of these sinusoidal modulations of PC$_2$ is much longer than dynamical time (t$_{dyn} \sim$ P = 0.65 days) and sublimation times \citep[$t_{sub} \sim$1 day][]{rappaport} for this system.

\begin{figure*}[!hbtp]
\begin{centering}
\includegraphics[width=0.49\textwidth]{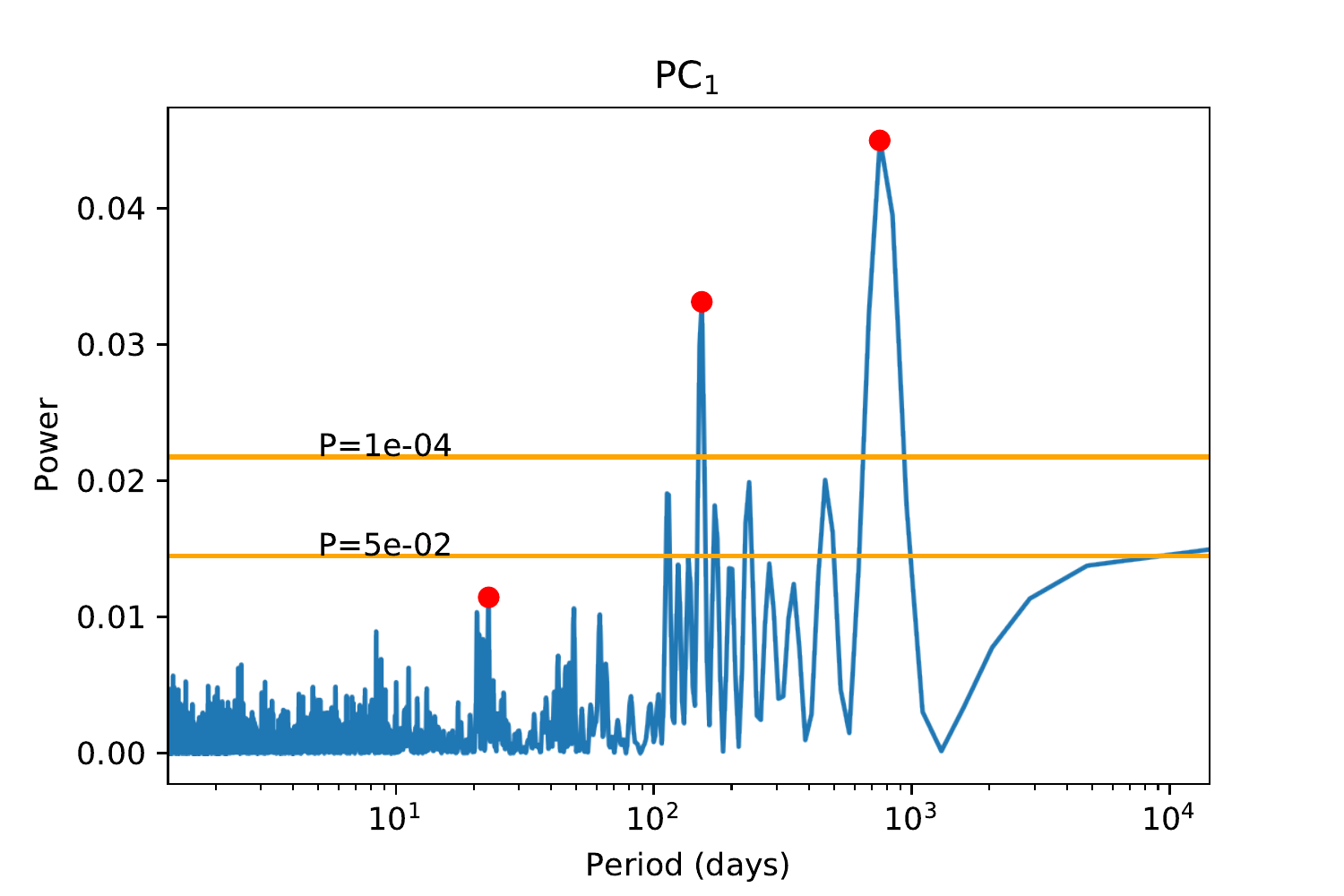}
\includegraphics[width=0.49\textwidth]{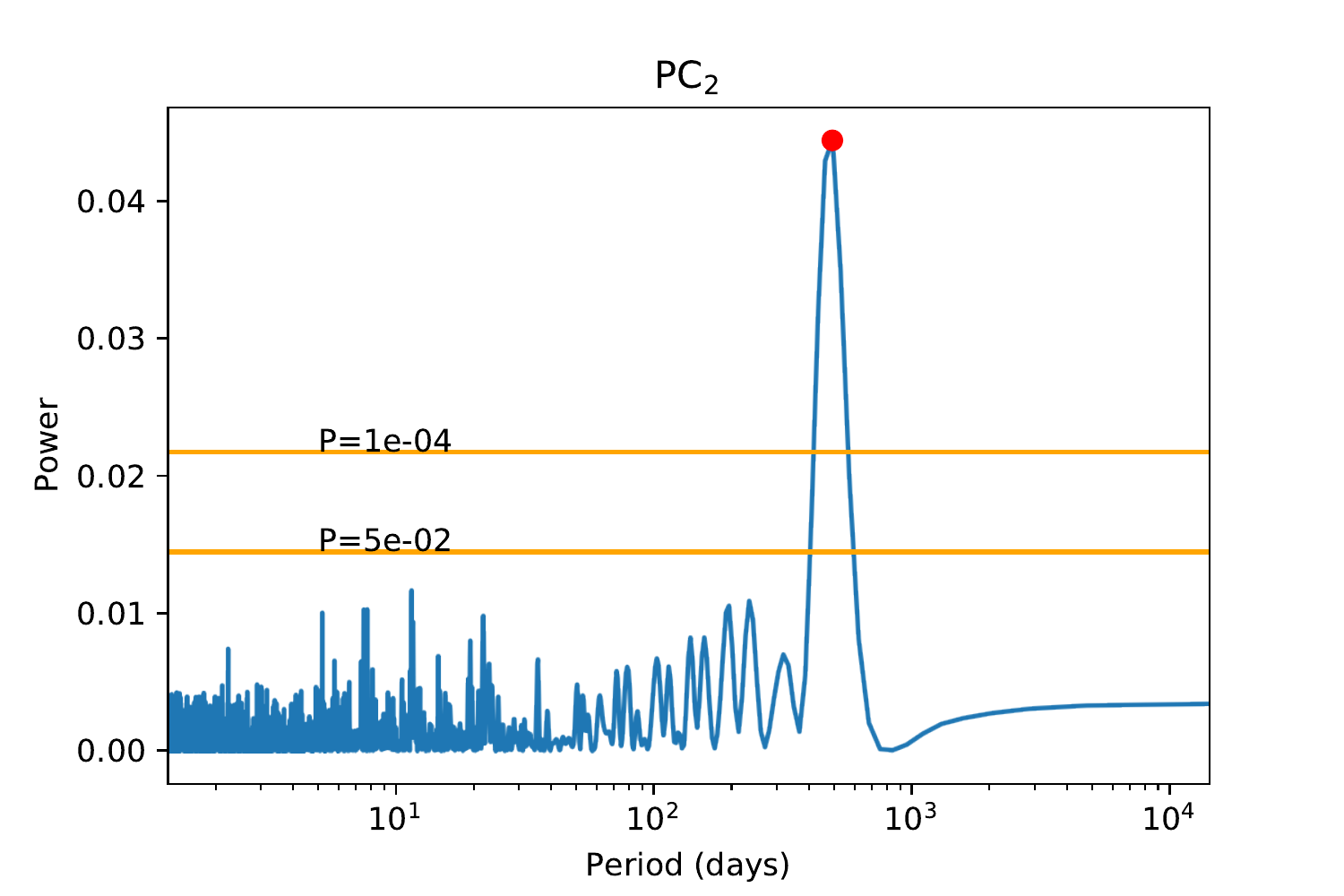}
\caption{Periodograms of the principal component time series shown in Figure \ref{fig:tserPC}, with peaks marked with red circles.
\deleted{The first peak at 0.654 days for both principal components is the orbital period, which is the sampling cadence of the principal components' time series.}
PC$_1$ \deleted{also} peaks at 22.9, 153 and 750 days.
The 22.9 day peak is consistent with \citet{kawahara2013starspots}, who notice that this period is co-incident with the stellar rotation period.
The second principal components shows one dominant peak at 491 days, \deleted{in addition to the 0.654 orbital period,} indicating possible long term evolution of the dust particle sizes.
\added{False alarm levels with probabilities of $5 \times 10^{-2}$ and $10^{-4}$ are shown as orange horizontal lines.}}\label{fig:ampPeriodogram}
\end{centering}
\end{figure*}

In Figure \ref{fig:ampPeriodogram} we show the periodogram of the principal components.
\added{We also show the formal false alarm levels of $5 \times 10^{-2}$ and $10^{-4}$ for the periodograms.}
\deleted{The peaks at 0.654 days in both the first and second principal components are the same as the orbital period, which determines the sampling cadence of the time series of amplitudes and principal components.}
The first component, corresponding to the transit depth, has peaks at 22.9, 153 and 750 days.
The 22.9 day peak is consistent with the \citet{kawahara2013starspots} and \citet{croll2015starspots} that shows amplitudes are anti-correlated with stellar flux.
As suggested in \citet{kawahara2013starspots}, the anti-correlation points to a physical mechanism for the disintegration - such as magnetic activity or high energy flux associated with spots causing disintegration on the planet.
The second principal component's periodogram has a strong peak at 491 days.
This singly-peaked periodogram \deleted{(other than the 0.654 day sampling period)} implies that PC$_2$ is nearly sinusoidal.
As shown in Figure \ref{fig:pcaVectorsComp}, the second principal component resembles models with larger dust grains, so the dust grain sizes could be evolving on long-term 491-day timescales.
This is much longer than the dynamical and sublimation times of the system, so there is no obvious mechanism for the dust grain size evolution.
\added{However, this timescale is close to the overturning timescale for a magma pool \citep{kite2016atmInteriorExchange}, which could control the source of metal rich vapor that creates grains.} 
It is also possible that red noise such as 1/$f$ noise contributes to these long term oscillations.

\added{
We note that the long-term evolution of the transit depths and shape may be similar to the trends observed on WD1145+017 \citep{gary2017photometricObsWD1145p017,rappaport2018opticalActivityWD1145p017}.
WD1145+017 also changes its transit activity on $\sim$ month long timescales.
These trends in transit activity are explained by either random collisions of asteroids or variable dust emission of asteroids.

We suspect that sublimating asteroids around WD1145+017, \sha\ and K2-22 b have long term behavior determined by surface features.
If alternating layers of rocky material have different composition, sublimation temperatures or albedos, they could provide a natural way to regulate the sublimation rate of the surface.
Alternatively, the chance alignment of starspots with the planet could regulate the planet disintegration \citep{kawahara2013starspots} and create periodic behaviors depending on the alignment of stellar rotation with the planet's position within its orbit.
\sha\ exhibits large stochastic variance that occasionally shuts off during quiescent intervals.
The transit activity level of WD1145+017's exhibits more complicated time dependence because there are many separate planetesimals or recently broken-up fragments that can serve as sources of dust \citep{vanderburg2015wdDisintegrating}.
}

\subsection{Statistical Analysis of Ground-based Photometry}\label{sec:statistics}
\begin{figure*}[!hbtp]
\begin{centering}
\includegraphics[width=0.49\textwidth]{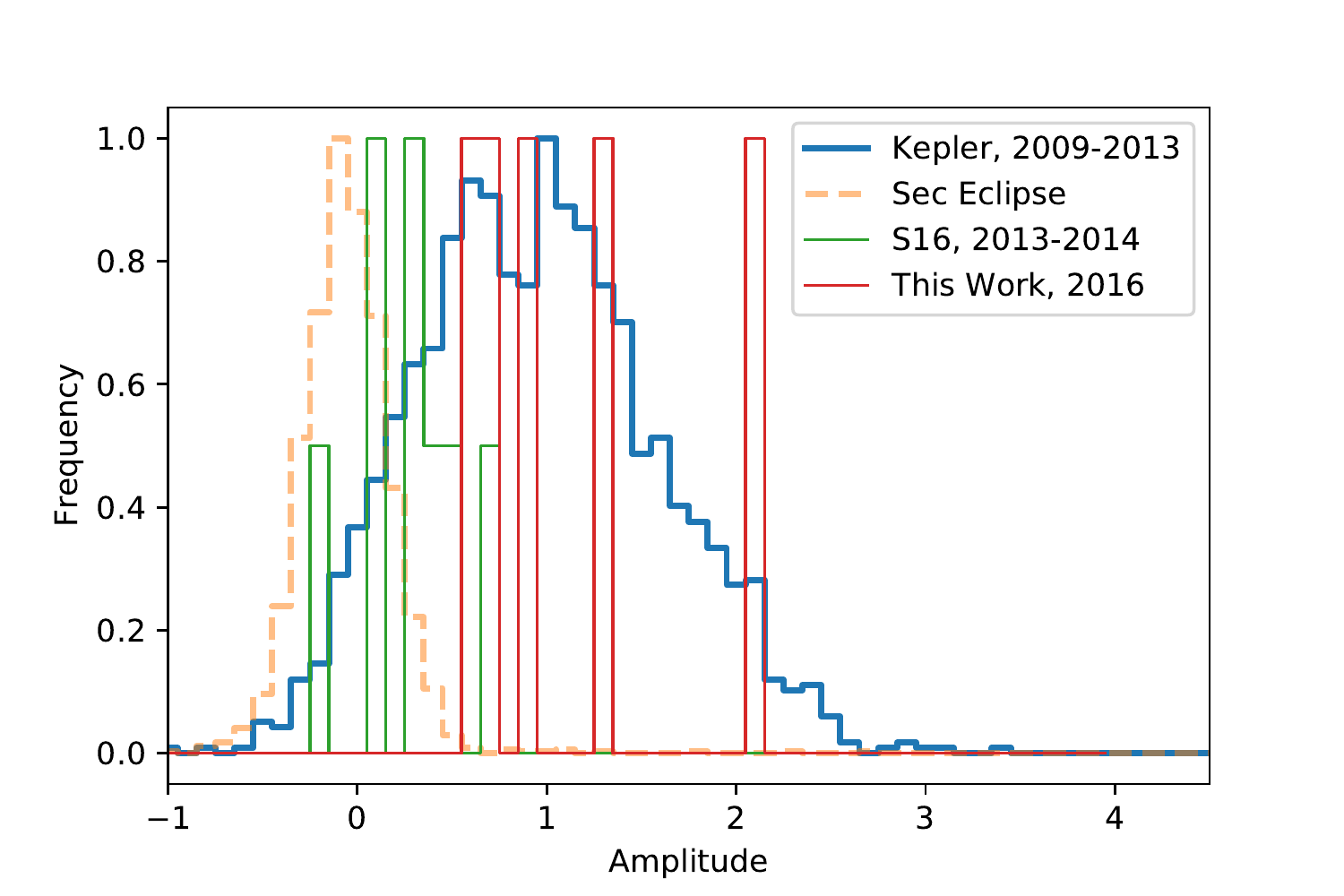}
\caption{Normalized frequency of 4 different amplitude distributions: 1) all of the \kepler\ Long Cadence data during transit from 2009 to 2013, 2) the secondary eclipse amplitudes from the \kepler\ spacecraft over this same period, 3) the 8 night ground-based IRTF transit campaign in the summers of 2013 and 2014 from \citet{schlawin2016kic1255} (S16) and 4) the 5 night Kuiper telescope transit campaign in the summer of 2016.
The \citet{schlawin2016kic1255} (S16) events give a KS-test $p$ value of 0.001 when comparing to the \kepler\ data.
In contrast, the 2016 photometric results in this work are consistent with the \kepler\ Long Cadence data, giving a  $p$ value of 0.5.
This indicates that the transit depths are back to normal and there is no slowdown of disintegration activity over time.}\label{fig:histoPhot}
\end{centering}
\end{figure*}

We calculate and evaluate the \kepler\ transit amplitudes to compare to ground-based results and to understand the long-term behavior of \sha's disintegration.
We take the histogram of transit amplitudes as measured by the \kepler\ light curves (shown in Figure \ref{fig:histoPhot}) and compare these to the ground-based photometry from \citet{schlawin2016kic1255} and this work.
The histogram of secondary eclipse amplitudes from the \kepler\ light curves is also plotted in Figure \ref{fig:histoPhot}, which shows the spread due to photon noise.
The intrinsic astrophysical variation of the planet disintegration is convolved by this photometric scatter in the data.

We compare the distribution of amplitudes from \citet{schlawin2016kic1255} from the \kepler\ long cadence data from 2009 to 2013.
A Kolmogorov-Smirnov (KS) test gives a $p$ value of 0.001 of the null hypothesis that the \citet{schlawin2016kic1255} results are drawn from the same distribution as the \kepler\ Long Cadence data.
This indicated a possible slowdown of disintegration activity.
However, the distribution from this work taken in 2016 shows consistency with the \kepler\ histogram, with the KS test $p$ value of 0.5 of the null hypothesis. 
We therefore conclude that the disintegration activity is stochastic with no long-term slowdown measurable from ground-based photometry.

\section{Stellar Characterization}\label{sec:stellarCharacterization}
\subsection{Previous Observations}

\begin{deluxetable*}{lcrrr}
\tablecaption{Observational estimates of the stellar parameters of \shStar, adapted from \citet{vanlieshout2016kic1255}}
\tablewidth{0pt}
\tablehead{
\colhead{Reference} &
\colhead{$ T_\mathrm{eff,*} $} &
\colhead{$\log(g)$} &
\colhead{Evolutionary Status} &
\colhead{Method} \\
 & (K) & $\log$(cm s$^{-2}$) & & \\
 }
\startdata
  \citet{brown2011kic} & 4400 $\pm$ 200   & 4.6 $\pm$ 0.5    & main-sequence star & photometry \\
  \citet{rappaport}       & 4300 $\pm$ 250    &                            & main-sequence star & low-resolution spectroscopy \\
  \citet{kawahara2013starspots} & 4950 $\pm$ 70 & 3.9 $\pm$ 0.2 & sub-giant          & high-resolution spectroscopy \\
  \citet{huber2014kicprop} & 4550$^{+140}_{-131} $ & 4.622$^{+0.043}_{-0.036} $& main-sequence star & photometry \\
  \citet{morton2016falsePos} & 4677 		& 4.61		& main-sequence star	& photometry \\
  \citet{vanlieshout2016kic1255} & 		& $\gtrsim 4.4$	& main-sequence star	& transit light curve \\
  This work - Specmatch	& 4440 $\pm$70 	& 4.63 $\pm$ 0.12 & main-sequence star & high-resolution spectroscopy\\
  This work - BOSZ		& 4500	& 4.5		& main-sequence star & high-resolution spectroscopy\\
\enddata
\tablenotetext{}{While photometry and low-resolution spectroscopy indicated surface gravity of a main sequence star, the high resolution analysis from \citet{kawahara2013starspots} indicated a gravity consistent with a sub-giant star.}\label{tab:stellObsParams}
\end{deluxetable*}

The host star \shStar\ has been characterized by both photometry and spectroscopy to better understand the system and parameters of the disintegrating planet and debris.
Table \ref{tab:stellObsParams} shows the summary of observations and analyses compiled in \citet{vanlieshout2016kic1255} for the stellar properties, re-produced here with additional results added from \citet{morton2016falsePos} and this work.
In the photometric analysis and low resolution spectroscopy, the spectra are consistent with a 4500 K main sequence K-type star.
High resolution spectra, however, indicated a higher temperature 4900 K lower log(g) sub-giant star \citep{kawahara2013starspots}.
\citet{vanlieshout2016kic1255} analyzed the light curve with a dust model to put constraints on the semi-major axis in terms of stellar radii, which can be combined with Kepler's third law to calculate the stellar density \citep{seager2003uniqueSolution}.
\citet{vanlieshout2016kic1255} find that the stellar density is only consistent with a main-sequence star with log(g) $\gtrsim 4.4 \log($cm s$^{-2})$.
\citet{vanlieshout2016kic1255} suggest an intriguing possibility that the high resolution spectrum of \shStar\ has contamination from dusty debris from the planet and thus appears as a sub-giant star.
If true, high resolution spectra would be \added{a} valuable tool to study the composition and dynamics of the material escaping \sha.
In this section, we analyze archival high resolution spectra to put new constraints on the stellar temperature and surface gravity.

\subsection{Subaru HDS Spectra}\label{sec:SubaruDescrip}

We evaluate the stellar parameters by examining the archival spectra taken with the High Definition Spectrograph (HDS) on the Subaru telescope \citep{noguchi2002hds}.
An observing log of the 3 nights is summarized in Table \ref{tab:specObs}.
There is one set of data taken on 2013 June 22 (UT) discussed in \citet{kawahara2013starspots}, including 2 exposures each 2700 seconds in duration.
For these 2013 exposures, the instrument was configured to use an image slicer unit \#2, which has a resolving power R$\approx$80,000.
Additionally, there are two sets of observations from 2015 Aug 28 (UT) and 2015 Aug 29 (UT) that were taken to test if the planet could be on a high eccentricity grazing orbit \citep{masuda2018rvKIC1255}, with exposure durations of 2400 seconds each.
The instrument was configured with a 0.3 mm (0.6\arcsec) wide and 30 mm long (60\arcsec) slit, which has a resolving power of R $\approx$60,000.
The second set of observations happened after the reaction wheel failure on 2013 May 11\footnote{\url{https://www.nasa.gov/mission_pages/kepler/news/keplerm-20130521.html}} that ended photometry of the main \kepler\ field, so no simultaneous \kepler\ photometry was available with the high dispersion spectroscopy to assess disintegration activity preceding or following the spectra.
None of the observations were timed during a primary transit (phase = 0.0) or secondary eclipse (phase = 0.5 for a circular orbit).
Future high resolution spectroscopy at an orbital phase of 0.0 may be useful for searching for gaseous planetary debris in absorption against the star.

\begin{deluxetable*}{cccccccc}
\tablecaption{List of Spectroscopic Observations}
\tablefontsize{\scriptsize}
\tablehead{\colhead{Exposure Number} & \colhead{ExpTime} & \colhead{UT Date} & \colhead{UT Start} & \colhead{Image Slicer} & \colhead{Slit Width} & \colhead{Orbital Phase} & \colhead{Resolving Power}\\
	& (s)	& & & & (mm) & & R}
\startdata
94085 & 2700 & 2013 Jun 22 & 12:32 & 2 & 2.0 & 0.621 & 80,000 \\
94087 & 2700 & 2013 Jun 22 & 13:18 & 2 & 2.0 & 0.670 & 80,000 \\
111345 & 2400 & 2015 Aug 28 & 05:37 & N & 0.3 & 0.667 & 60,000 \\
111349 & 2303 & 2015 Aug 28 & 06:19 & N & 0.3 & 0.712 & 60,000 \\
111355 & 2400 & 2015 Aug 28 & 07:33 & N & 0.3 & 0.790 & 60,000 \\
111359 & 2400 & 2015 Aug 28 & 08:15 & N & 0.3 & 0.835 & 60,000 \\
111373 & 2400 & 2015 Aug 28 & 09:17 & N & 0.3 & 0.901 & 60,000 \\
111525 & 2400 & 2015 Aug 29 & 05:34 & N & 0.3 & 0.194 & 60,000 \\
111529 & 2400 & 2015 Aug 29 & 06:17 & N & 0.3 & 0.240 & 60,000 \\
111533 & 2400 & 2015 Aug 29 & 07:00 & N & 0.3 & 0.285 & 60,000 \\
111549 & 2400 & 2015 Aug 29 & 08:31 & N & 0.3 & 0.382 & 60,000 \\
111553 & 2400 & 2015 Aug 29 & 09:14 & N & 0.3 & 0.428 & 60,000 \\
\enddata
\tablenotetext{}{Observing log of the Subaru HDS spectroscopic observations.}\label{tab:specObs}
\end{deluxetable*}

\subsection{Data Reduction}
For the 2015 data, we use the same spectra, as in \citet{masuda2018rvKIC1255}.
All telluric emission features and outliers are masked in this analysis to remove these components.
Time-variable outliers, such as cosmic rays, are removed by measuring deviations from the median spectrum for the night.
The night-sky emissions from OH and O$_2$ are removed by using the \citet{osterbrock1996lineAtlas} night sky atlas.
All OH and O$_2$ lines are removed regardless if they are visible in the spectrum of \shStar.
The wavelength of each spectrum is shifted to correct for the observatory's motion about the Solar Systems barycenter as well as the systematic -36.3 km/s velocity \citep{croll2014}.
We combined the normalized spectra from UT dates 2015 Aug 28 and 2015 Aug 29 with a median in the time direction, which is more robust to cosmic rays than an average.

For the 2013 data, we follow the \texttt{iraf} reduction techniques from the instrument manual V2.0.0  \footnote{\url{https://www.subarutelescope.org/Observing/Instruments/HDS/}}.
Some modifications to the manual were made, including a more aggressive -50/+50 lower and upper window for bad pixel identification with \texttt{mkbadpx} on the red CCD.
We applied bad pixel masks and bias files to each CCD (red and blue sides) separately.
We performed wavelength calibration with a Thorium Argon reference spectrum.
The final wavelength calibration fit was performed with a chebyshev polynomial, 5th order in the x direction and 3th order in the y direction.
This polynomial fit has a RMS error of 0.002$\AA$.

\begin{figure*}[!hbtp]
\begin{centering}
\includegraphics[width=1.0\textwidth]{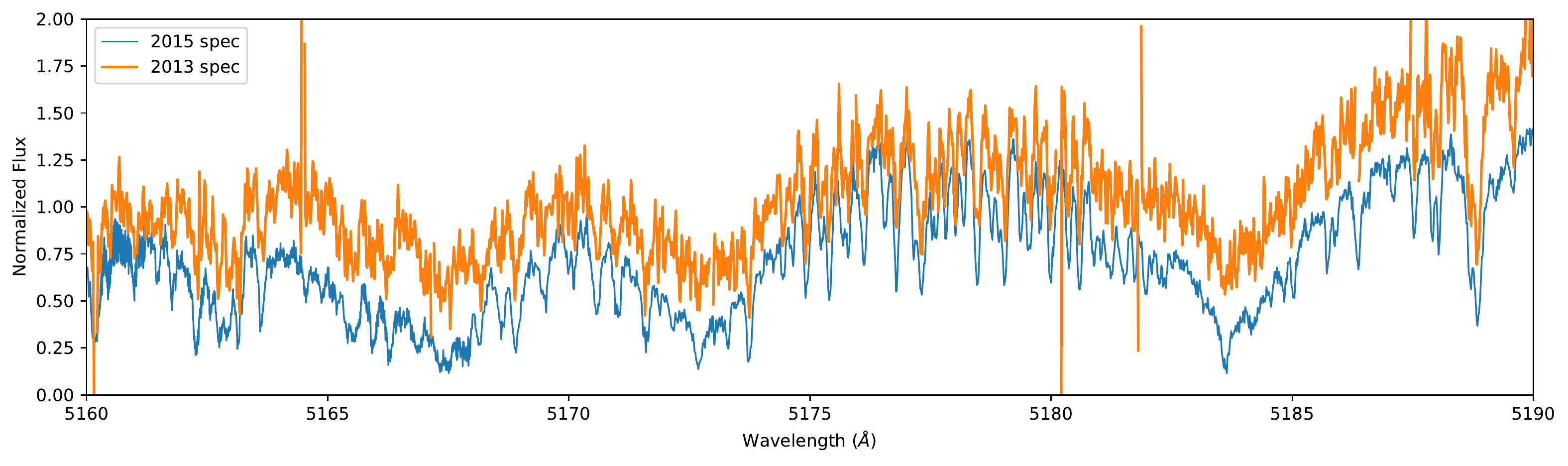}
\caption{Comparison between the 2013 spectrum and 2015 median spectrum of \shStar\ near the Mg I triplet. The two are consistent within the larger noise of the 2013 data.}\label{fig:spec2013vs2015}
\end{centering}
\end{figure*}

We compare the 2013 and 2015 data to search for any variability in the absorption lines, as seen in Figure \ref{fig:spec2013vs2015}.
Absorption line variability could be caused by disintegration activity of the planet that escapes as dusty material and then sublimates into gas.
This gas absorption could provide kinematic information on escaping material provided that the sublimated gas has sufficient optical depth.
We find no statistically significant differences between the 2013 and 2015 spectra, as shown in Figure \ref{fig:spec2013vs2015}.
We then proceed with the 2015 data to constrain stellar models since it has 6.7 hours of total exposure time (compared to 1.5 hours in 2013).

\subsection{Comparison to BOSZ}
We first evaluate stellar parameters by comparing the high resolving power ($R=100,000$) BOSZ grids of ATLAS-APOGEE ATLAS9 models \citep{bohlin2017bosz} to the median spectrum from 2015.
Because the discrepancies in the literature are mostly differences in the surface gravity (see Table \ref{tab:stellObsParams}), we study the gravity-sensitive Mg triplet near 517 nm.
The $R=100,000$ models are convolved with a Gaussian kernel that has a standard deviation $\sigma_K = \lambda/161,000 \approx 0.03 \AA$ to best match the data.
This was \replaced{smaller}{slightly broader} than we expected from combining inverse resolving powers in quadrature, which would imply that an \replaced{$R=75,000$}{$R_{expected} = 75,000$} kernel \added{($\sigma_{expected} = \lambda/176,600$)} convolved with an $R=100,000$ spectrum would result in a $R=60,000$ spectrum.
\added{We suspect that the non-Gaussian shape of the instrumental line broadening is the cause of this discrepancy.}

\begin{figure*}[!hbtp]
\begin{centering}
\includegraphics[width=0.85\textwidth]{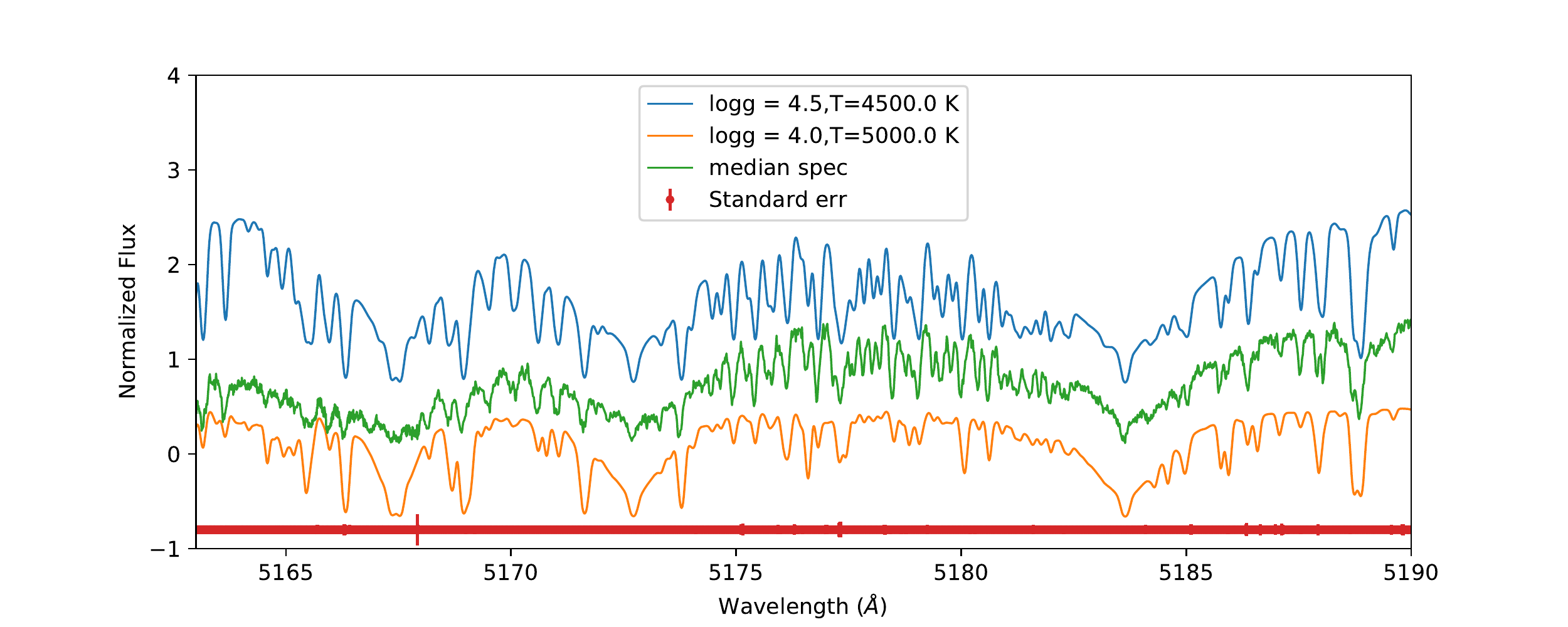}
\caption{Median spectrum of \shStar\ \citep{masuda2018rvKIC1255} near the Mg triplet (green) compared to 4500 K BOSZ models with two different log(g)s (blue and orange). The standard error in the mean is shown at an arbitrary Y location as a red line.}\label{fig:mgTripletBOSZ}
\end{centering}
\end{figure*}

We start by comparing the measured spectrum with two BOSZ models in Figure \ref{fig:mgTripletBOSZ}.
Here, the two BOSZ models are broadly representative of the two types of results in Table \ref{tab:stellObsParams} (log(g)=4.5 and log(g)=4.0).
The higher gravity log(g)=4.5 model appears to match the data better in terms of line ratios but does not explain all of the line depths in the spectrum.

\begin{figure*}[!hbtp]
\begin{centering}
\includegraphics[width=0.45\textwidth]{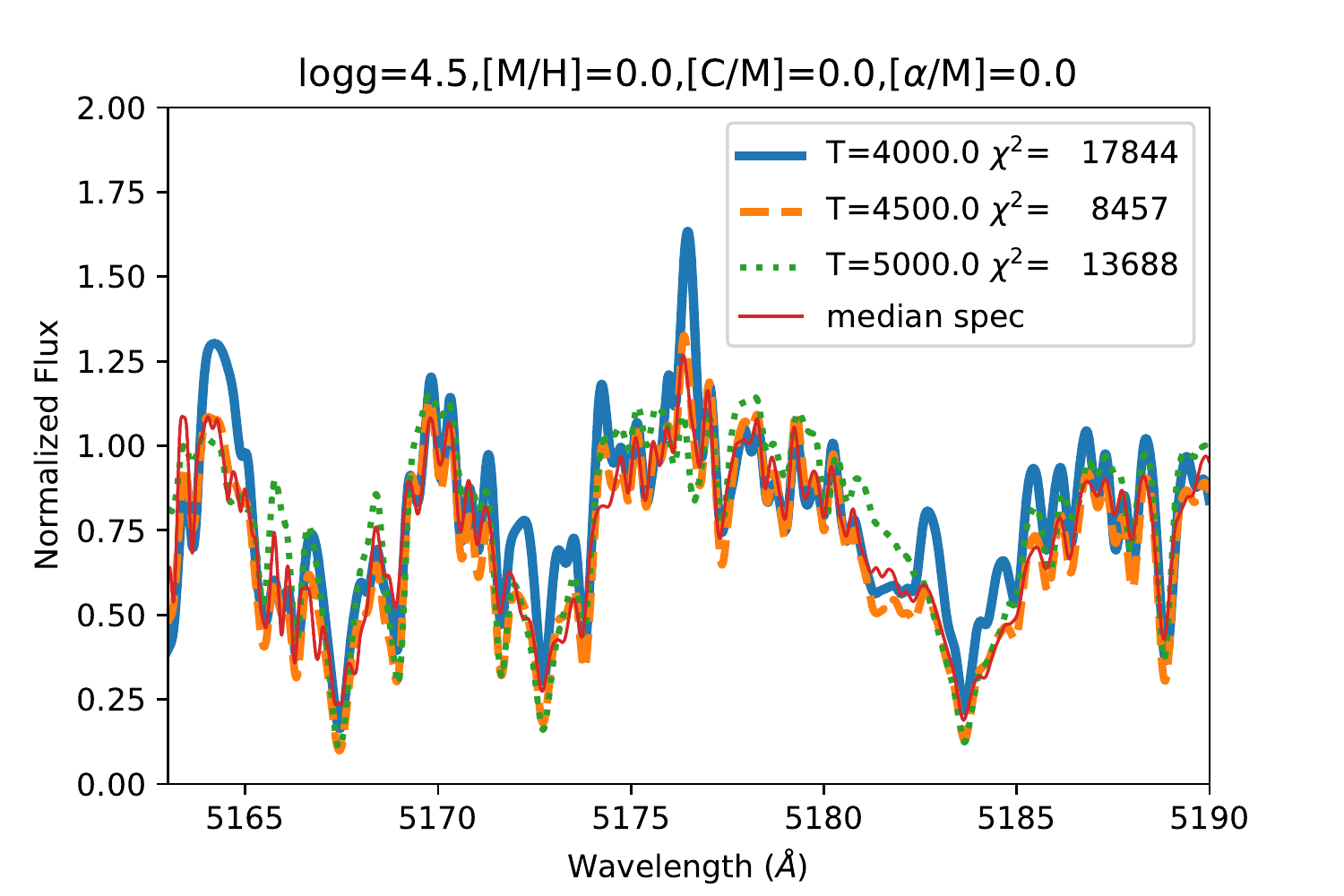}
\includegraphics[width=0.45\textwidth]{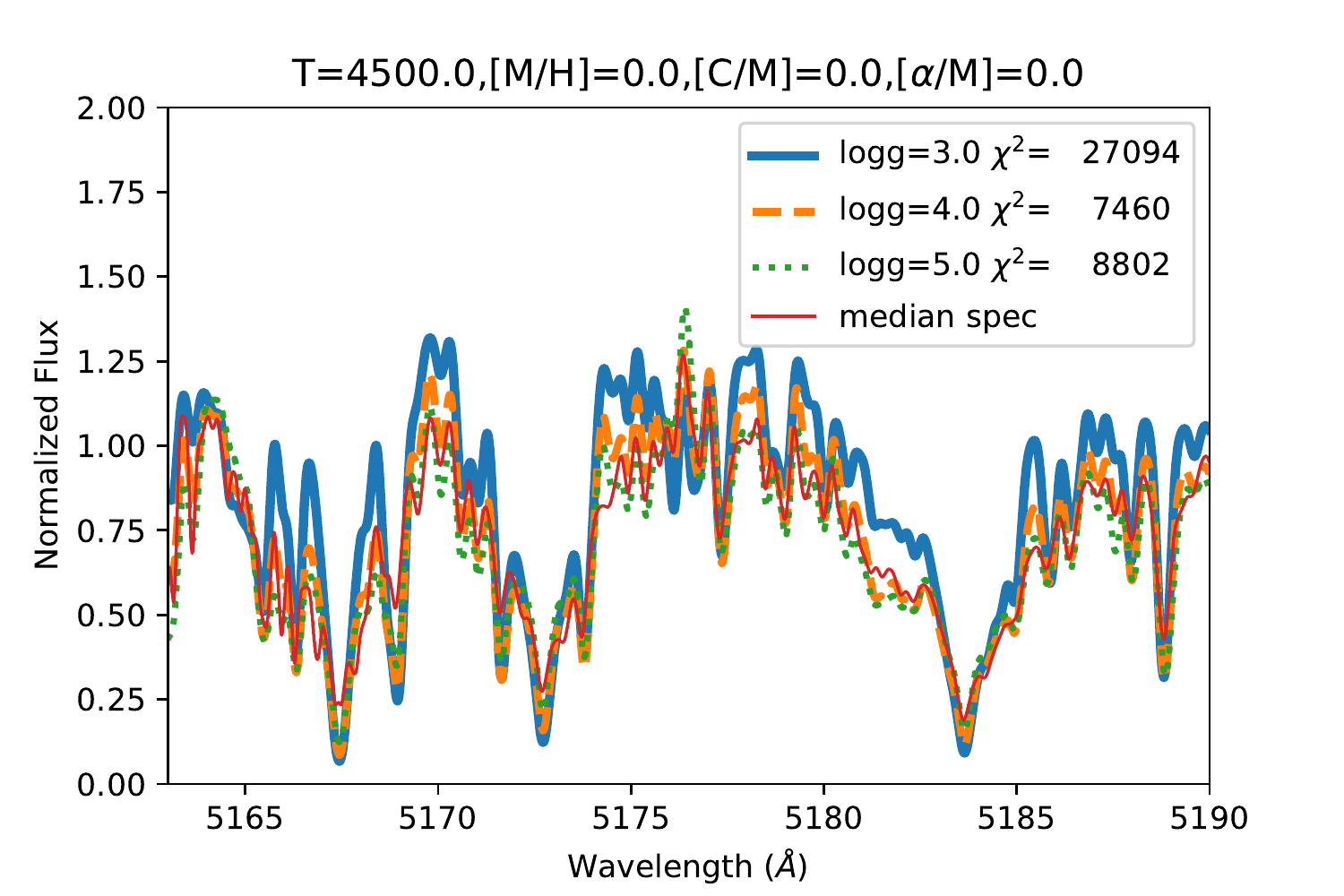}
\includegraphics[width=0.45\textwidth]{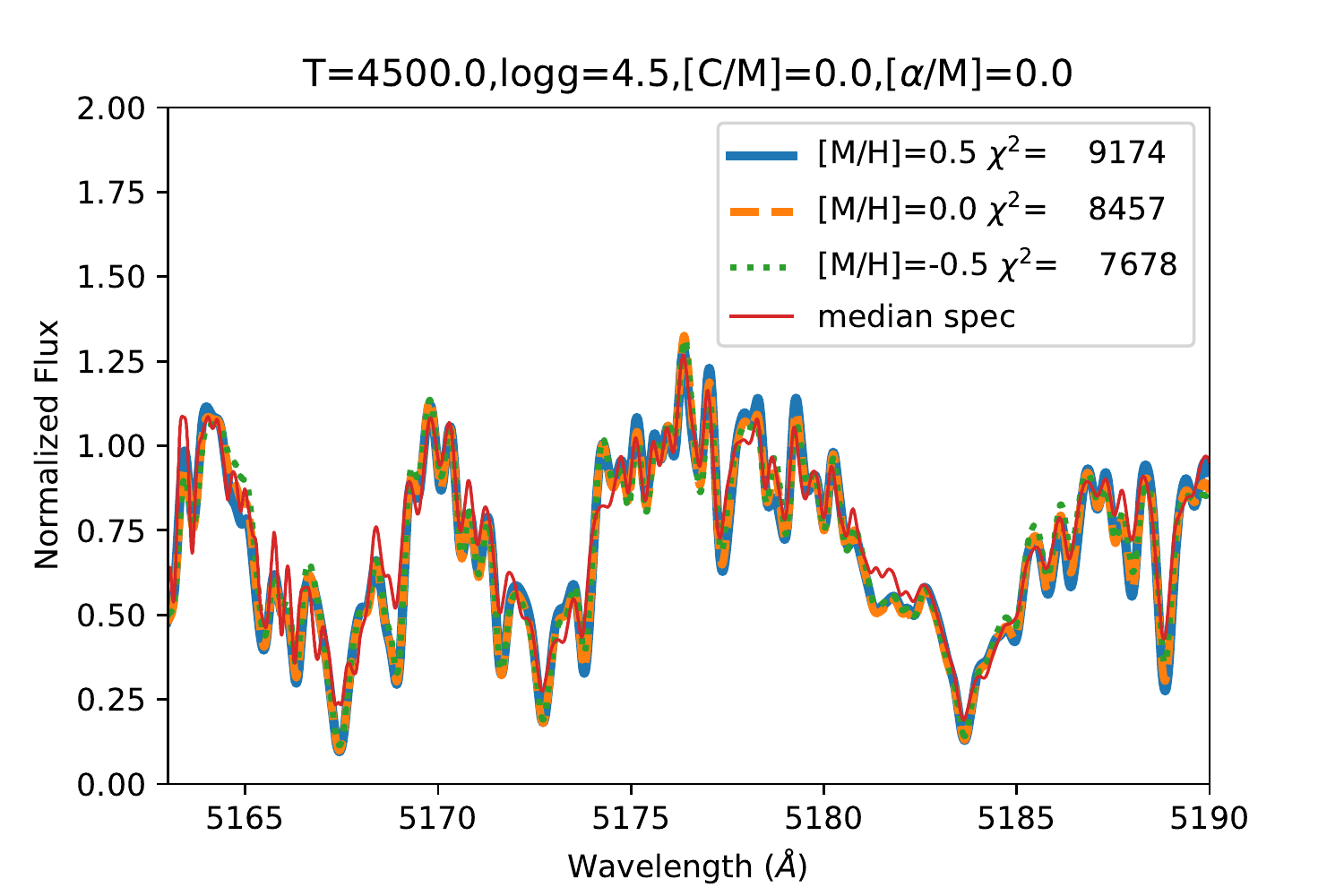}
\includegraphics[width=0.45\textwidth]{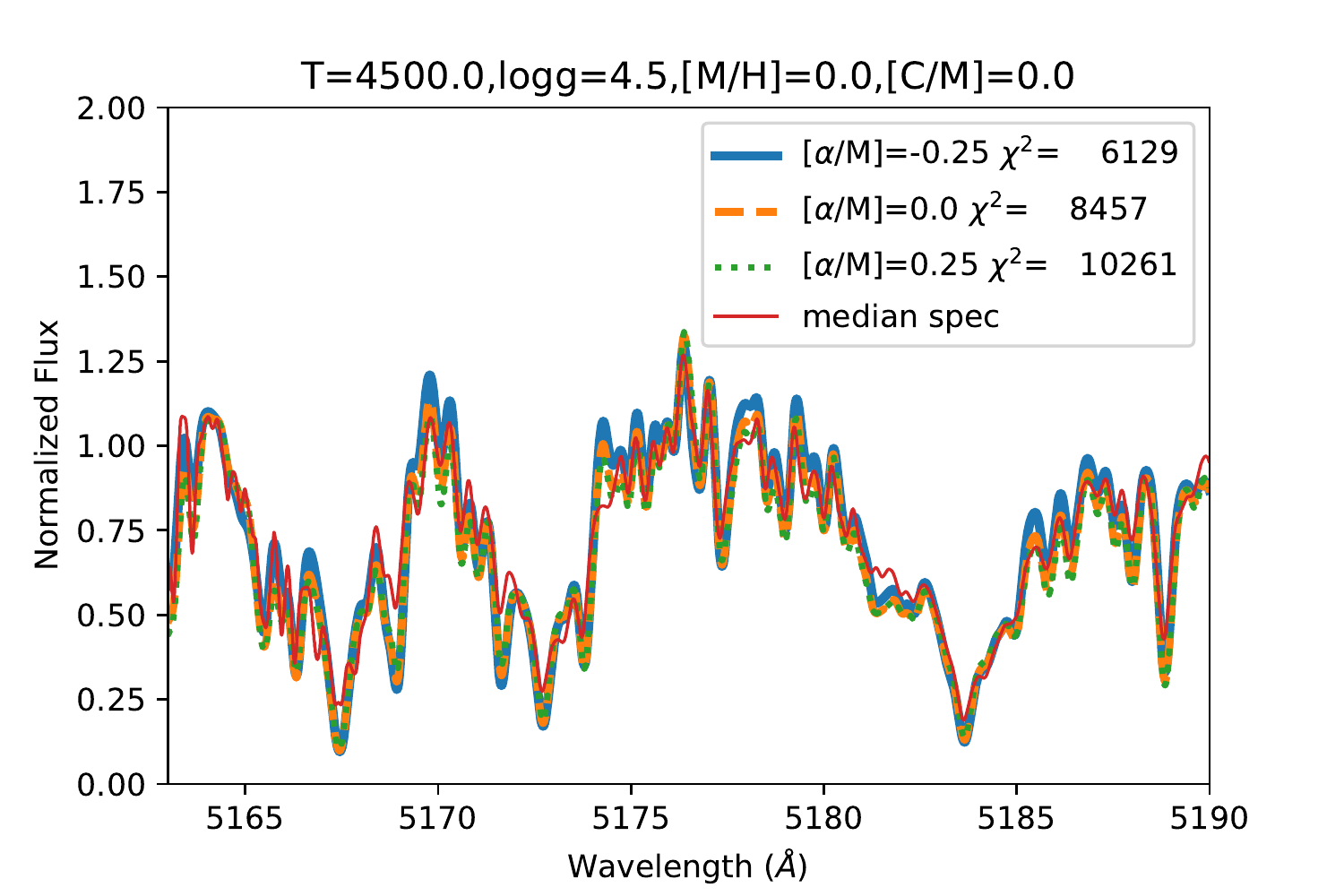}
\includegraphics[width=0.45\textwidth]{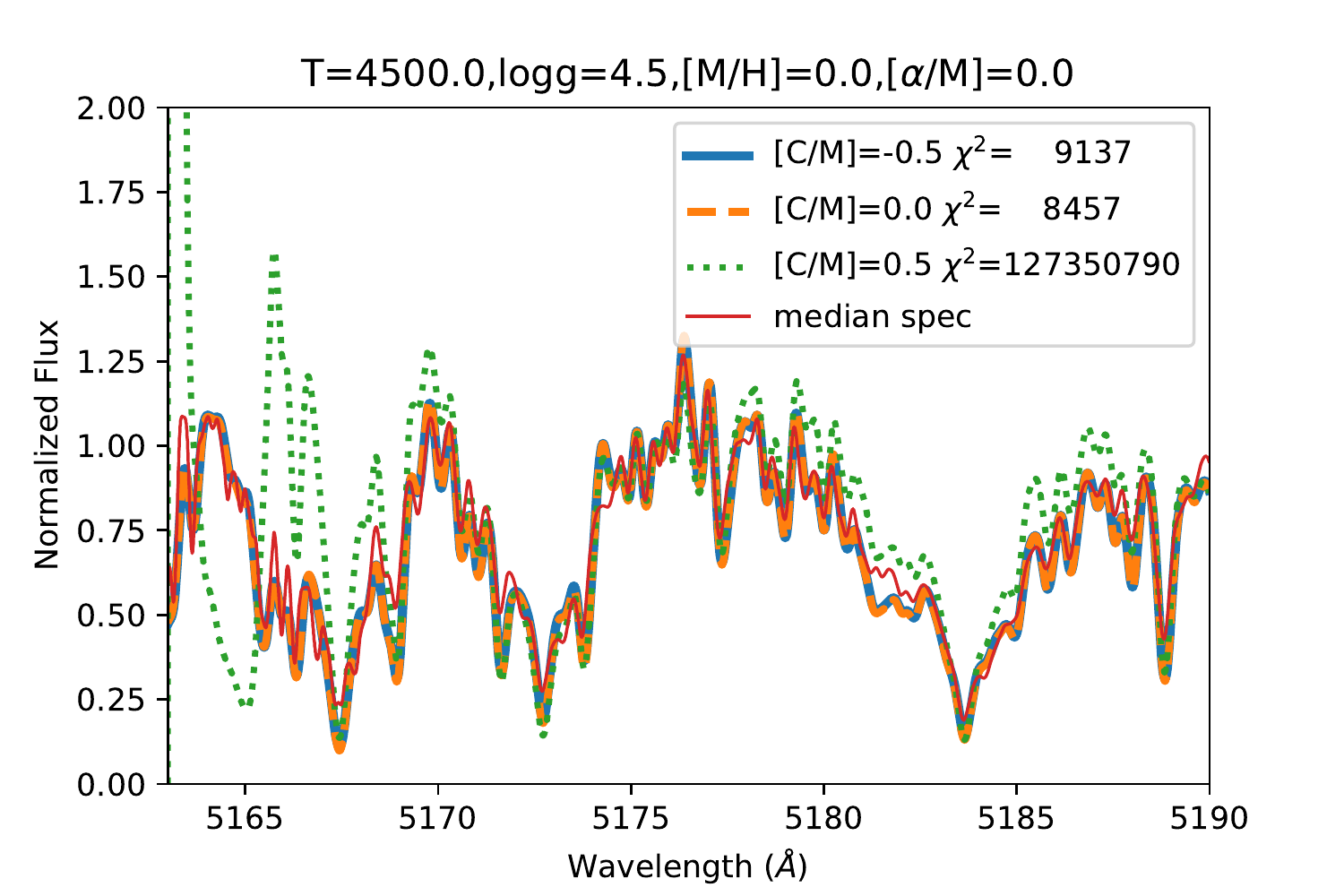}
\includegraphics[width=0.45\textwidth]{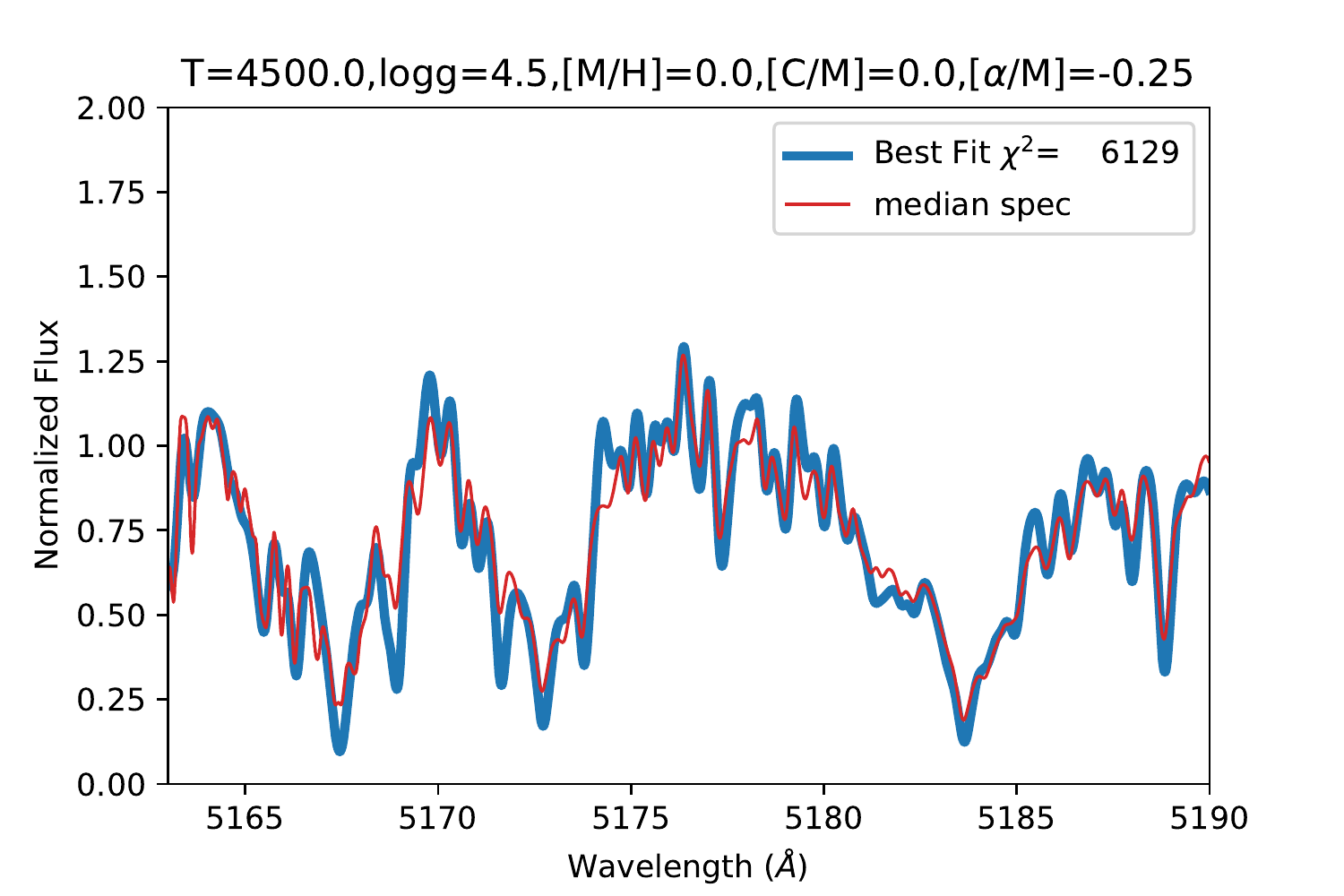}
\caption{The BOSZ models are explored over the parameter space to search for a best fit as well as visualize how much each parameter affects the spectrum.
The bottom right plot shows the best-fit BOSZ model, but it still under-predicts and over-predicts some line strengths.}\label{fig:boszModelParamsMedianSpec}
\end{centering}
\end{figure*}

We explore the 5 parameters in the BOSZ models: T$_\mathrm{eff}$, log(g), [M/H], [$\alpha$/M] and [C/M] to constrain these parameters in \shStar.
The sensitivity to each parameter for the BOSZ models are shown in Figure \ref{fig:boszModelParamsMedianSpec}.
For each parameter, we change the models by about 2 steps in the BOSZ grid.
For a quantitative measure of the goodness-of-fit, we calculate a $\chi^2$ over the Mg triplet lines from 5160 to 5190 $\AA$.
The errors are estimated from the standard deviation of the spectra over the night.

After initial constraints on the models from the sensitivity to parameters search, we downloaded a grid of BOSZ models with effective temperatures ranging from 4250 K to 4750 K, Log(g) from 4.0 to 4.5, [M/H] from 0.0 to 0.5, [C/M]=0.0 and [$\alpha$/M] = -0.25 to 0.0.
We found the model with the  $\chi^2$ statistic over the Mg I triplet lines from 5160 to 5190 $\AA$.
The minimum $\chi^2$ model for this entire grid is the same as shown in Figure \ref{fig:boszModelParamsMedianSpec} on the Bottom Right: T$_\mathrm{eff}$=4500 K, Log(g)=4.5, [M/H]=0.0, [C/M]=0.0, $\alpha=-0.25$.
This is similar to the log(g)$\approx$4.6, T$_\mathrm{eff}$=4500 stellar parameters derived from photometry and low resolution spectroscopy listed in Table \ref{tab:stellObsParams}.

\subsection{SpecMatch-Emp}\label{sec:SpecMatch-Emp}

We also use the open-source \texttt{SpecMatch-Emp} tool \citep{yee2017specMatch} to estimate the stellar parameters using the Subaru spectrum.
\texttt{SpecMatch-Emp} uses a library of Keck HIRES spectra to match the spectrum and derive stellar parameters.
We cut the default HIRES library and Subaru spectra to the wavelengths from 5000~$\AA$ to 5900~$\AA$, which includes the gravity-sensitive Mg I triplet.

\begin{figure*}[!hbtp]
\begin{centering}
\includegraphics[width=0.8\textwidth]{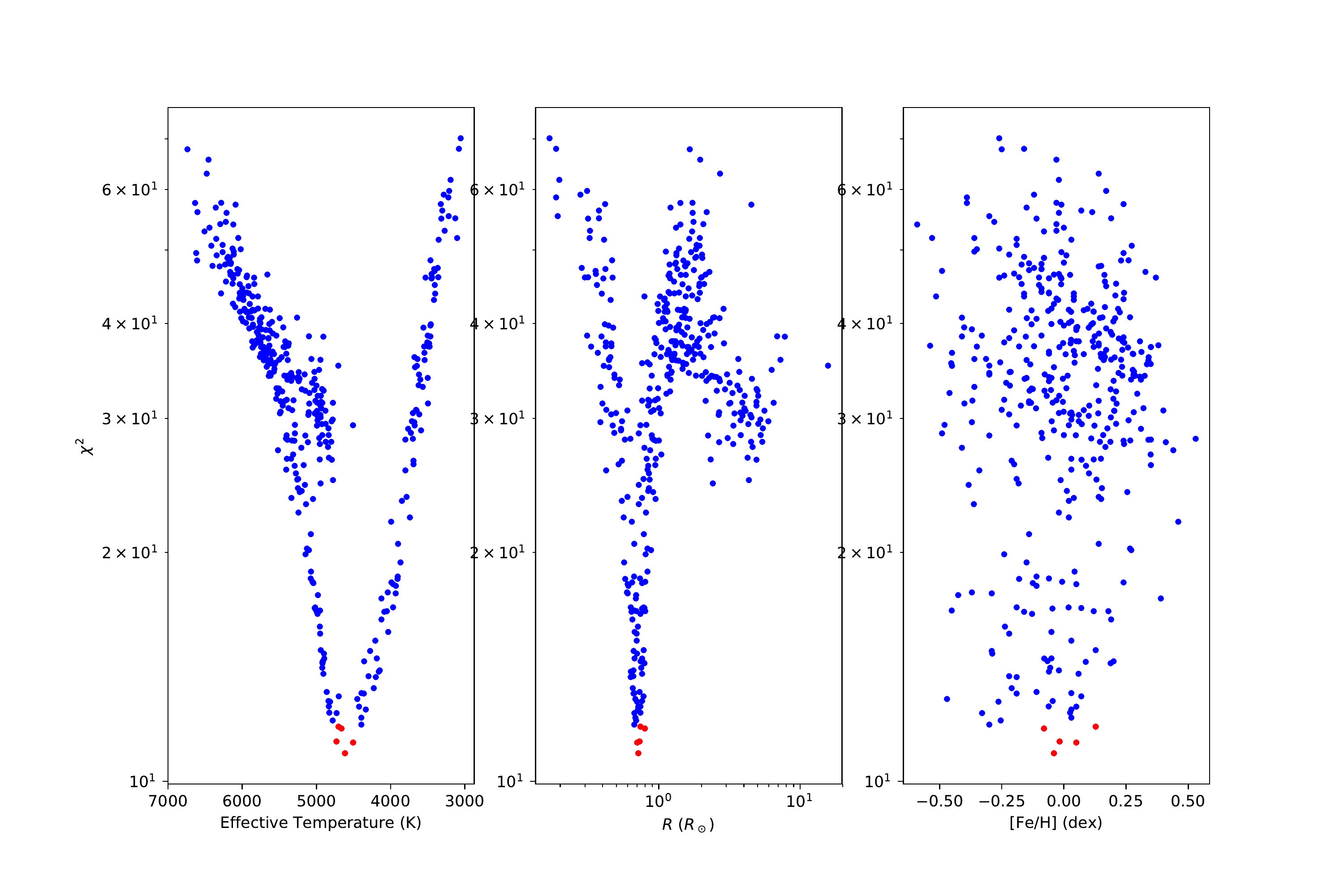}
\caption{$\chi^2$ as a function of stellar parameters found by \texttt{SpecMatch-Emp}.
The effective temperature and radii show clear global minima.}\label{fig:SpecMatch-Emp}
\end{centering}
\end{figure*}

\begin{figure*}[!hbtp]
\begin{centering}
\includegraphics[width=0.8\textwidth]{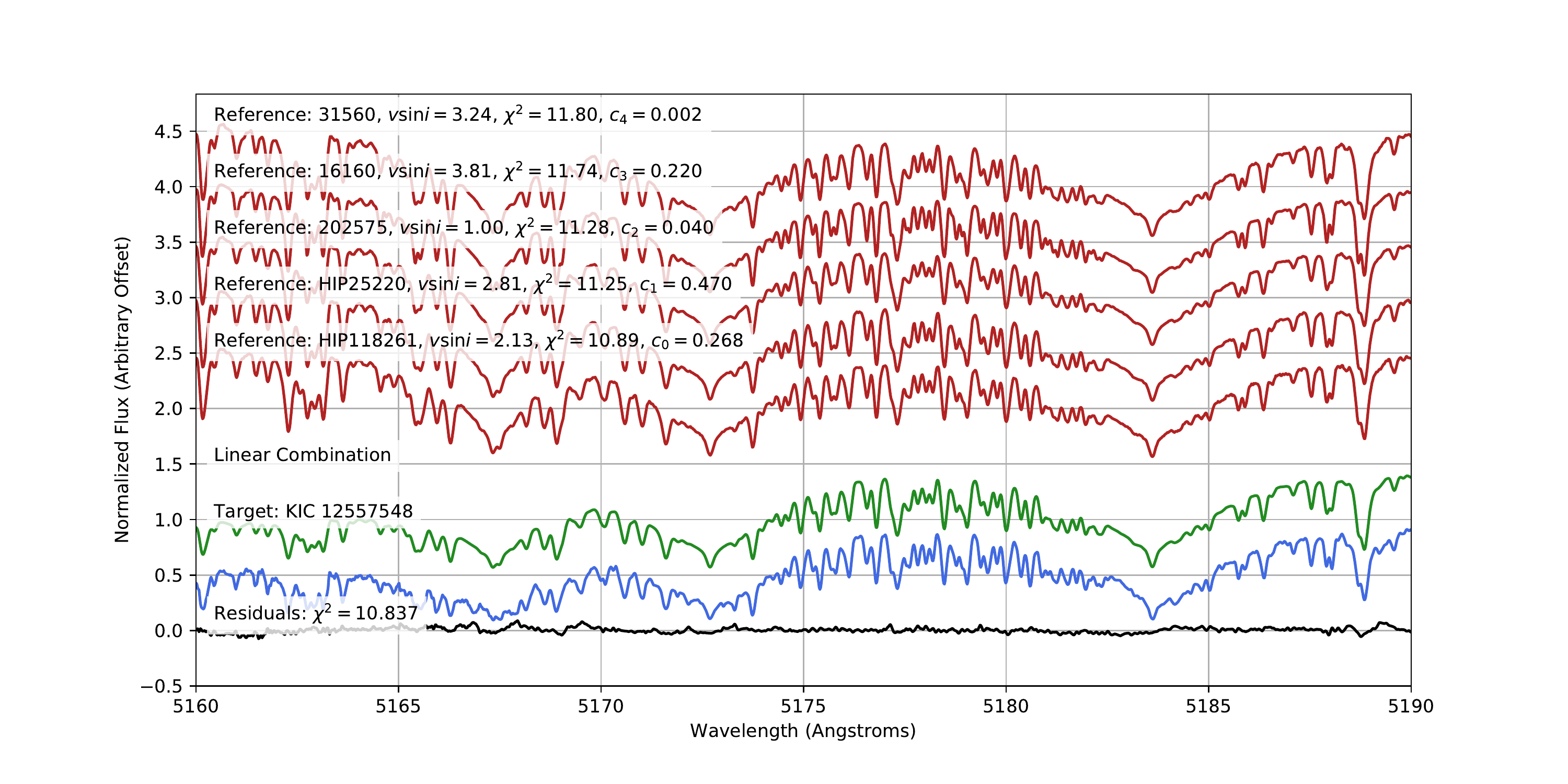}
\caption{Linear combination of templates that best matches KIC 1255 near the Mg I triplet.}\label{fig:SpecMatch-EmpComb}
\end{centering}
\end{figure*}

Figure \ref{fig:SpecMatch-Emp} shows the chi-squared differences between the median Subaru spectrum and the models.
The effective temperature and radius parameters show clear global $\chi^2$ minima for the HIRES spectral libraries.
However, the metallicity is less well constrained.
The resulting best-fit parameters from \texttt{SpecMatch-Emp} are T$_\mathrm{eff}$=4440 K $\pm$ 70 K, [Fe/H]=-0.08$\pm$0.09, log(g)=4.63$\pm$0.12, R=0.69 $\pm$ 0.10 R$_\odot$ and Mass=0.70 $\pm 0.08~$M$_\odot$.
Figure \ref{fig:SpecMatch-EmpComb} shows the linear combination of spectra that best-fit \shStar\ from \texttt{SpecMatch-Emp} over a subset of the wavelengths near the Mg I triplet.

We summarize these \texttt{SpecMatch-Emp} and BOSZ results in Table \ref{tab:stellObsParams}, which both favor a main-sequence log(g) $\approx$ 4.6 star.
Our analysis of the median 2015 spectrum therefore indicates that the stellar spectrum shows no strong contamination from planet disintegration activity \added{at the level where it would affect the derived stellar parameters}.
Unfortunately, this makes diagnosing the gaseous material evaporating from sublimated dust grains challenging to detect.
A brighter transiting disintegrating system may reveal itself more strongly, if discovered in the TESS field.

We posit that the initial analysis of the 2013 Subaru spectrum \citep{kawahara2013starspots} that found a temperature and log(g) of 4950 K and 3.9 log(cm/s$^{-2}$) was affected by insufficient signal to noise in the spectrum.
The 2013 data has a 30\% smaller count rate (in e$^-$/s) than the 2015 data likely because the 2013 weather and guiding were sub-optimal.
On top of that, the total exposure time for the 2013 observations was only 1.5 hours.
The consequence of the increased noise in the 2013 spectra affected the interpretation of this high resolution spectrum.
For example, the SNR of the Na D lines was $\sim$ 30, which is near the threshold needed for the \citet{takeda2005specFGKdwarfs} equivalent width method employed in the initial analysis.
We therefore conclude that the noise in the spectrum affected the stellar parameters rather than gas absorption from sublimating dust grains.
\added{
\subsection{Spectral Energy Distribution and Parallax}\label{sec:SED}

We also confirm our finding that the star is a main sequence using the Gaia DR2 parallax of 1.617 $\pm$ 0.030 mas \citep{gaia2016mission,gaia2018dr2,luri2018parallaxes}.
We follow the methodology from \citet{stassun2017gaiaRadiiMasses}, which is to estimate the bolometric flux from the spectral energy distribution and invert the Stefan-Boltman law to solve for the radius.
For the measured spectral energy distribution, we use the VizieR photometry service and fit this with a representative 4600 K \citet{castelli2004models} model and a (E-B)=0.05 diffuse Milky Way extinction model \citep{cardelli1989}.
We use the \texttt{pysynphot} tool \citep{lim2015pysynphot} to calculate the model and extinction.
We note that the exact stellar model used is less important than finding a function to fit the Spectral Energy Distribution.
When inverting the Stefan-Boltzman law, we assume an effective temperature of 4440 $\pm$ 70 K from our spectral fitting in Section \ref{sec:SpecMatch-Emp} and a photometric error of 5\%, assumed to be dominated by photometric systematics.
Using this method we find a radius of 0.72 $\pm$ 0.03 R$_\odot$, which is consistent with a main sequence star.
}

\added{
\section{Stellar Activity}\label{sec:activity}

As discussed in Section \ref{sec:intro}, \citet{kawahara2013starspots} first noted an anti-correlation between the transit depths of \sha\ and the flux of its host star.
This correlation implies that the planet disintegration is regulated by the high energy radiation or magnetic energy associated with star spots and that the planet's passage over the spot can increase the disintegration activity.
\citet{croll2015starspots} confirm this anti-correlation but offer a different explanation: occulted dark spots on the stellar surface can decrease the transit depth because the planetary debris extincts less total starlight.
If there is a localized grouping of spots that the planet's transit chord passes through, this could create a positive correlation of transit depth with stellar flux.
\citet{croll2015starspots} note that the spots could have a complicated distribution and that a separate set of spots on the other side of the star could dominate the overall light curve rotational modulation but be at a different latitude from the transit chord, thus creating an anti-correlation between transit depth and stellar flux.
\citet{croll2015starspots} calculate that un-occulted spots have a minimal impact on the transit depth variations.

Recently, \citet{rackham2018transitSourceEffect} have studied in detail the effect of un-occulted spots on the transit depths of exoplanets orbiting M stars, the so-called transit source light effect.
Using a suite of model rotating photospheres with active regions added successively at random positions, they explored the dependence of observed brightness variations with spot and faculae covering fractions.
One commonly held assumption is that there is a linear relation between the amplitude of variability and the spot covering fraction.
Instead, \citet{rackham2018transitSourceEffect} find that the amplitude of variability is proportional to $\sqrt{f_S}$, where $f_S$ is the spot covering fraction.
They also find that large ranges of $f_{S}$ are consistent with a given amplitude of variability, leading to a correspondingly large range of transit depth changes possible for a given stellar variability amplitude.

Extending the analysis to spectral types F5V to K9V, which display larger spot contrasts \citep{berdyugina2005spots} and lower typical rotational variabilities than M dwarfs \citep{mcquillan2014rotationPeriodsAutoC}, \citet{rackham2018activityFGKsubmitted} find the observed variabilities of FGK dwarfs point to relatively low spot coverages and transit depth biases.
Their scaling relations suggest that the observed amplitude variations of 0.8\% \citep{kawahara2013starspots} for a K4 dwarf correspond to $f_{S}=3^{+5}_{-2}\%$, indicating that transit depths of exoplanets can be biased by $3^{+3}_{-2}\%$ at optical wavelengths depending on the distribution and sizes of spots and that the transit chord is completely devoid of spots.

High values of $f_S$ are possible for longitudinally symmetric distributions of active regions, such as a latitudinal band or polar spot.
Polar spots are indeed observed on rapidly rotating stars via Doppler imaging \citep{strassmeier2009starspots}; however, one would be unlikely on \sha, given its relatively long rotation period.
Additionally, the large physical size of \sha's tail, as indicated by a transit duration longer than the time to cross the star, means that spots are likely to be crossed and this diminishes the transit source effect further.
In this light, \sha's $\sim$30\% observed transit depth variation at a $\sim$ 22 day period would require a pathological distribution of magnetically active regions to be explained by un-occulted spots.
If the spot distribution is polar or distributed in a latitudinal band, this would produce a bias at all epochs, unlike the time-dependent transit depth variations observed in \sha.
Therefore, the conclusion from \citet{croll2015starspots} that un-occulted spots are unlikely to cause the correlation between transit depth and stellar flux is likely correct even when considering a more realistic scaling relation between spot covering fraction and rotational modulation amplitude.
}

\section{Conclusions}\label{sec:conclusions}

We obtained ground-based R band photometry from the Kuiper 61-inch telescope to follow up a possible slowdown in disintegration activity observed in 2013 and 2014.
\citet{schlawin2016kic1255} found shallow transits that were all below the \kepler\ observatory average in August-September 2013 and August-September 2014.
One possibility was a long term evolution of the disintegration activity, such as a reduction of available dusty material of the planet or a slowdown in the disintegration mechanism.
Our transit depths from June-July 2016 are instead consistent with the depths from the \kepler\ observatory from 2009 to 2013, with a KS-test between the two distributions resulting in a p-value of 0.5 that depths are drawn from the same distribution.
This indicates that the shallow transit depths observed in August-September 2013 and August-September 2014 likely fell into the 14-36 day long quiescent intervals of disintegration activity as observed in \kepler\ photometry \citep{kawahara2013starspots,vanWerkhoven2014,croll2015starspots}.

We re-analyzed the existing \kepler\ photometry with PCA to better understand the time dependence and statistics of the light curves.
We found that the first eigenvector corresponds to the overall transit depth whereas the second eigenvector corresponds to models of forward-scattering from  large-sized dust grains.
The first principal component tracks closely with the transit depths and exhibits modulations at periods of 22.9, 153 and 750 days.
The second principal component is nearly sinusoidal with a 491 day period, indicating possible long term evolution of the dust particle sizes.
The 22.9 day period matches the rotation period of the host star, which has previously been used to tie disintegration to stellar activity \citep{kawahara2013starspots} or occultations of star spots \citep{croll2015starspots}.
The remaining periodicities are longer than the dynamical and sublimation times of the planet and grains, so a different mechanism would be required to modulate the activity over these timescales.

\citet{vanlieshout2016kic1255} compiled the stellar parameters reported in the literature via photometric and spectroscopic methods and note that the high resolution spectrum show that \shStar\ has the gravity of a sub-giant, whereas all other methods find that it is a main sequence star.
One exciting possibility suggested in \citet{vanlieshout2016kic1255} is that the planet disintegration affects the spectral lines of the star and affects the high resolution spectrum from \citet{kawahara2013starspots}.
We explore this possibility by examining archival Subaru HDS spectra of \shStar.
We find that the high resolution spectra are consistent with previous photometry and spectroscopy and that the host star is on the main sequence with T$_\mathrm{eff}$=4440 $\pm$ 70 K, log(g) = 4.63 $\pm$ 0.12, [Fe/H]=-0.08$\pm$0.09, \replaced{M}{R}=0.69$\pm$0.10 R$_\odot$ and M=0.70 $\pm$0.08 M$_\odot$.
Therefore, the optical depth of any gas sublimating \replaced{of}{off} dust grains is too small to affect \added{the parameters derived from} the high resolution spectrum.
We argue that low signal to noise in the initial Subaru HDS spectra affected previous interpretations of the stellar parameters.
\added{We also confirm the main-sequence radius of the star using the Gaia DR2 parallax.
We examine the effect of un-occulted spots on the transit depth behavior in light of recent research by \citet{rackham2018transitSourceEffect} that star spot coverage can be underestimated by a linear scaling with stellar flux variability amplitude.
Even when accounting for the larger coverage of starspots by this effect, we confirm the conclusion from \citet{croll2015starspots} that {\it un-occulted} spots are unlikely to explain the correlation between transit depth and stellar flux.
It is still possible that \sha's disintegration is regulated by stellar activity or that {\it occulted} spots account for this correlation.}

\section{Acknowledgements}
The authors thank to Kento Masuda for sharing the 2015 Subaru spectrum of \shStar\ and giving helpful comments on this work.
Funding for E Schlawin is provided by NASA Goddard Spaceflight Center.
\replaced{The authors thank Benjamin Rackham for providing an estimate of the transit light source effect for a K4 V star and for helpful comments from Saul Rappaport.}{The authors thank Saul Rappaport for helpful comments.} 
\added{Thanks to Ben Weiner for some useful discussion on high resolution spectroscopy.
This work was supported by the Japan Society for Promotion of Science (JSPS) KAKENHI Grant Number JP16K17660.
Support for this work was provided by NASA through Hubble Fellowship grant HST-HF2-51399.001 awarded by the Space Telescope Science Institute, which is operated by the Association of Universities for Research in Astronomy, Inc., for NASA, under contract NAS5-26555.}
\added{This research has made use of the VizieR catalogue access tool, CDS, Strasbourg, France. The original description of the VizieR service was published in \citet{ochsenbein2000vizieR}.
This work has made use of data from the European Space Agency (ESA) mission {\it Gaia} (\url{https://www.cosmos.esa.int/gaia}), processed by the {\it Gaia} Data Processing and Analysis Consortium (DPAC,\url{https://www.cosmos.esa.int/web/gaia/dpac/consortium}).
Funding for the DPAC has been provided by national institutions, in particular the institutions participating in the {\it Gaia} Multilateral Agreement.
This publication makes use of data products from the Wide-field Infrared Survey Explorer, which is a joint project of the University of California, Los Angeles, and the Jet Propulsion Laboratory/California Institute of Technology, funded by the National Aeronautics and Space Administration.}
The results reported herein benefited from collaborations and/or information exchange within NASA?s Nexus for Exoplanet System Science (NExSS) research coordination network sponsored by NASA?s Science Mission Directorate.

\acknowledgments




\facility{Subaru, SO:Kuiper, Kepler}

\software{
	\texttt{ccdproc} \citep{craig2015ccdproc},
	\texttt{astropy} \citep{astropy2013}, 
	\texttt{scikit-learn} \citep{pedregosa2011scikit-learn},
	\texttt{photutils v0.3} \citep{bradley2016photutilsv0p3},
	\texttt{iraf},
	\texttt{SpecMatch-Emp} \citep{yee2017specMatch},
         \texttt{emcee} \citep{foreman-mackey2013emcee}, 
         \added{ \texttt{pysynphot} \citep{lim2015pysynphot}}
          }

\bibliographystyle{apj}
\bibliography{ms}

\end{document}